\documentclass[11pt,aps,prl,onecolumn,superscriptaddress,groupedadress]{revtex4-2}

\usepackage{comment}
\usepackage{amsmath}
\usepackage[dvips]{graphicx}
\usepackage{epsfig}
\usepackage{tabularx}
\usepackage{lineno}
\usepackage{ulem}
\usepackage{xcolor}
\usepackage{soul}

\usepackage[pdfpagemode=UseNone,colorlinks=true,linkcolor=blue,citecolor=blue]{hyperref}

\newcommand{\ud}{\mathrm{d}}
\newcommand{\WX}{\Omega_{\scriptstyle{x}}}
\newcommand{\WY}{\Omega_{\scriptstyle{y}}}
\newcommand{\Wb}{\Omega_{\mathrm{b}}}
\newcommand{\WLO}{\Omega_{\scriptscriptstyle{LO}}}
\newcommand{\WALO}{\Omega_{\scriptscriptstyle{LO}}^{\scriptstyle{A}}}
\newcommand{\WBLO}{\Omega_{\scriptscriptstyle{LO}}^{\scriptstyle{B}}}
\newcommand{\aA}{\hat{a}_{\scriptscriptstyle{A}}}
\newcommand{\aB}{\hat{a}_{\scriptscriptstyle{B}}}
\newcommand{\ain}{a^{\scriptscriptstyle{in}}}
\newcommand{\aout}{a^{\scriptscriptstyle{out}}}

\newcommand{\matCA}{\mathbf{A}}
\newcommand{\matV}{\mathbf{V}}
\newcommand{\alphaLO}{\alpha_{\scriptscriptstyle{LO}}}
\newcommand{\gA}{g_{\scriptscriptstyle{A}}}
\newcommand{\gB}{g_{\scriptscriptstyle{B}}}
\newcommand{\xm}{x_{\scriptscriptstyle{meas}}}
\newcommand{\ppm}{p_{\scriptscriptstyle{meas}}}
\newcommand{\num}{\nu_{-}}
\newcommand{\Gammaxi}{\Gamma_{\scriptscriptstyle{\xi}}}
\newcommand{\xb}{x_{\mathrm{b}}}
\newcommand{\pb}{p_{\mathrm{b}}}

\begin{document}

\pagestyle{plain}
\date{\today}
\title{Stationary entanglement of a levitated oscillator with an optical field}

\author{Q. Deplano}
\thanks{These authors contributed equally}
\affiliation{Dipartimento di Fisica e Astronomia, Università di Firenze, via Sansone 1, I-50019 Sesto Fiorentino (FI), Italy}
\affiliation{INFN, Sezione di Firenze, via Sansone 1, I-50019 Sesto Fiorentino (FI), Italy}

\author{A. Pontin}
\thanks{These authors contributed equally}
\affiliation{CNR-INO, largo Enrico Fermi 6, I-50125 Firenze, Italy}

\author{F. Marino}
\affiliation{CNR-INO, largo Enrico Fermi 6, I-50125 Firenze, Italy}
\affiliation{INFN, Sezione di Firenze, via Sansone 1, I-50019 Sesto Fiorentino (FI), Italy}

\author{F. Marin}
\email[Electronic mail: ]{francesco.marin@unifi.it}
\affiliation{Dipartimento di Fisica e Astronomia, Università di Firenze, via Sansone 1, I-50019 Sesto Fiorentino (FI), Italy}
\affiliation{INFN, Sezione di Firenze, via Sansone 1, I-50019 Sesto Fiorentino (FI), Italy}
\affiliation{CNR-INO, largo Enrico Fermi 6, I-50125 Firenze, Italy}
\affiliation{European Laboratory for Non-Linear Spectroscopy (LENS), via Carrara 1, I-50019 Sesto Fiorentino (FI), Italy}

\begin{abstract}
Stationary entanglement between the motion of macroscopic objects and light is a long-standing goal of quantum optomechanics, with implications for both fundamental tests of quantum physics and emerging quantum technologies. We report the generation of quantum entanglement between the center-of-mass motion of a nanosphere levitated in an optical tweezer inside an optical cavity and the electromagnetic field. By heterodyne detection, we reconstruct the full set of optomechanical correlations and observe a violation of separability bounds between the mechanical motion and the quadratures of a propagating optical mode. This demonstrates the distribution of nonclassical correlations beyond the interaction region. The entanglement is generated at room temperature and remains robust over a broad range of parameters. Our results establish levitated optomechanical systems as a promising platform for continuous-variable quantum communication and for tests of macroscopic quantum physics.

\end{abstract}

\maketitle

Entanglement is one of the defining features of quantum mechanics \cite{Einstein1935,Schroedinger1935}, and demonstrating its control in macroscopic systems is a crucial goal with far-reaching implications for quantum technologies and for probing the foundations of quantum physics. Optomechanical platforms have enabled exquisite control over mechanical motion, leading to milestones such as the preparation of non-Gaussian states \cite{Arrangoiz-Arriola2019,Bild2023} and entangled mechanical oscillators \cite{Ockeloen-Korppi2018,Riedinger2018,Kotler2021,Mercier2021,Wollack2022}. Furthermore, pulsed optomechanical entanglement between a drum oscillator and a microwave \cite{Palomaki2013}, and between nano-oscillators and single photons \cite{Marinkovic2018}, has been demonstrated in ultra-cryogenic environments.

Optomechanical interaction can entangle distinct optical fields  even exploiting a classical mechanical system \cite{Barzanjeh2019,Chen2020}, but stationary entanglement between mechanical motion and propagating light has remained elusive, despite long-standing theoretical proposals \cite{Bose1997,Vitali2007_2,Paternostro2007,Genes2008a,Zippilli2015,Gut2020}. Such light-matter entanglement is of particular interest because it provides a natural interface between stationary and flying quantum degrees of freedom, enabling nonlocal correlations to be distributed across space. This capability is a key resource for quantum communication networks, where mechanical oscillators act as local quantum memories while optical fields distribute information between distant nodes~\cite{Mancini2003,Pirandola2006}.

Among optomechanical platforms, levitated systems have recently emerged as a particularly attractive approach \cite{Gonzales-Ballestero2021}, combining strong light–matter interactions \cite{Quidant2020,Ranfagni2021} with exceptional isolation from the environment~\cite{Toros2020A,Northup2024}. The center-of-mass motion of optically trapped dielectric nanospheres has already reached the one- and two-dimensional quantum regimes \cite{Delic2020,Magrini2021,Tebbenjohanns2021,Ranfagni2022,Piotrowski2023,Deplano2025,kamba2025}. 

In this work, we achieve a further milestone by demonstrating stationary optomechanical entanglement between the mechanical motion of a levitated nanosphere and a propagating optical field, evidenced by correlations violating separability bounds.

\begin{figure}[!htb]
    \centering
      \includegraphics[scale=0.3]{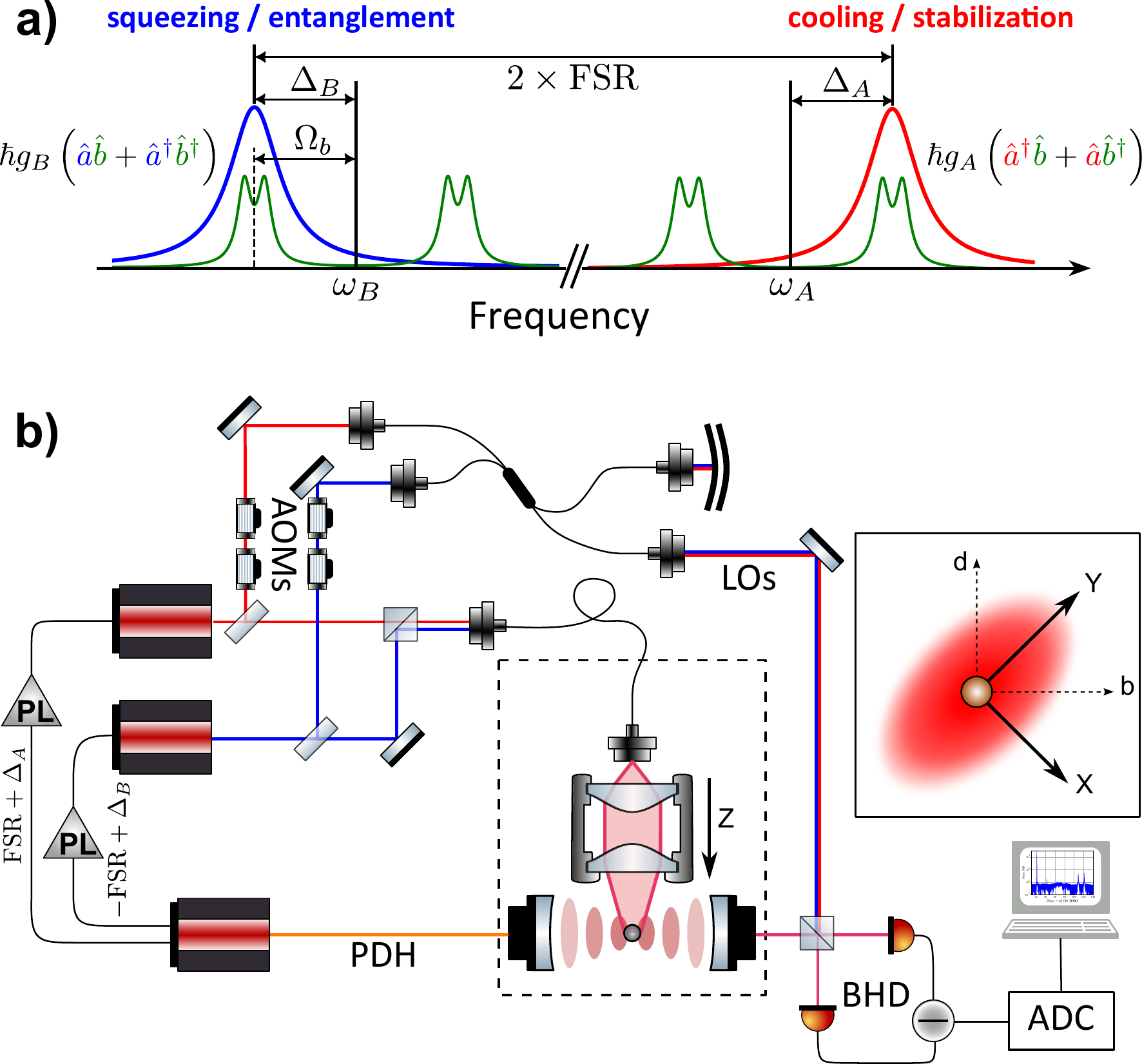}
    \caption[Fig1]{\textbf{Experimental configuration.} The motion of a levitated nanoparticle is coupled to the field of an optical cavity via coherent scattering. The confining potential is obtained by overlapping in the optical tweezer two fields, $A$ and $B$, with different frequencies. This enables simultaneous driving of the cavity on opposite sides of its resonances, with detunings $\Delta_A$ (red) and $\Delta_B$ (blue): the former cools and stabilizes the motion, while the latter generates entanglement between the mechanical and optical subsystems. \textbf{a)} The detuning of the $A$ and $B$ fields is set with respect to different longitudinal modes separated by two free spectral ranges (FSR) at a value close to the mechanical bright mode frequency $\Wb$.  \textbf{b)} This is achieved by phase locking (PL) both trapping lasers to an auxiliary laser, frequency stabilized to a cavity resonance using a Pound-Drever-Hall (PDH) locking scheme. The cavity output fields are measured in a heterodyne configuration using the same balanced detection (BHD). We use  two different local oscillators frequencies, detuned by 1.4~MHz and 2.0~MHz with respect to the corresponding trapping fields by means of acousto-optical modulators (AOM). The heterodyne signals are recorded after a high-resolution analog-to-digital converter (ADC). The inset depicts the beam shape of the tweezer (polarized along Y) in the transverse plane. The label "b" ("d") refers to "bright" ("dark") and corresponds to the direction along (orthogonal to) the cavity axis.
    Measurements are performed in UHV environment, at a pressure of $\simeq3.5\times10^{-8}$\,mbar.}
    \label{Fig1}
\end{figure}

\textit{Experimental procedure} - We load a 100 nm diameter silica sphere in an optical tweezer~\cite{Calamai2021} created by two laser fields (denoted $A$ and $B$) at 1064 nm, superposed in an optical fiber with identical linear polarization and a power ratio of 3:1. A slight ellipticity of the focused beam, with the major axis aligned along the polarization direction (see inset in Fig. \ref{Fig1}\,b)), split the transverse motion into two mechanical modes with eigenfrequencies $\WX/2\pi = 110.6 \,\mathrm{kHz}$ and $\WY/2\pi = 98.4 \,\mathrm{kHz}$. 
The nanosphere is placed at the center of an optical cavity whose axis is almost orthogonal to that of the tweezer. Both trapping lasers are phase locked to an auxiliary laser stabilized to a cavity resonance, enabling precise control of their detunings $\Delta_{A,B}$ from two cavity modes separated by twice the free spectral range ($\mathrm{FSR} = 3.07\,$GHz). The configuration is shown in Fig.~\ref{Fig1}\,a). Light scattered from the tweezer fields populates the two cavity modes, thereby coupling the particle motion to the intracavity fields via coherent scattering \cite{Vuletic2000,Windey2019,Delic2019A,Quidant2020}. Details of the experimental setup are provided in the Supplementary Materials \cite{SM1} and in Ref. \cite{Ranfagni2021}.

When the nanosphere is located near a node of the cavity standing wave, the dominant optomechanical coupling involves motion along the cavity axis (the bright mechanical mode \cite{Toros2020B,Toros2021,Borkje2023}, with eigenfrequency $\Wb \simeq (\WX+\WY)/2$). The red-detuned field $A$ directly cools this bright mode. Because the linear polarization is oriented at $\sim 45^{\circ}$ relative to the cavity axis, both transverse motional degrees of freedom are cooled. As a consequence, the direction orthogonal to the cavity axis (dark mode), even if it is not directly coupled to the cavity field, undergoes efficient sympathetic cooling \cite{Toros2021,Ranfagni2022,Deplano2025}. 
Operating in the resolved-sidebands regime, where the cavity decay rate $\kappa$ is smaller than $\Wb$, optimal cooling is achieved for detuning $-\Delta_A \simeq \Wb$. In this regime, the optomechanical interaction is well described by a beam-splitter Hamiltonian $\hbar \gA \left(\aA^{\dagger} \hat{b} + \aA \hat{b}^{\dagger} \right)$ ($\gA$ is the coupling rate, $\aA$ and $\hat{b}$ are the annihilation operators of the optical and mechanical modes, respectively). 
The field outgoing the cavity for each optical mode $\alpha=(A,B)$ is 
\begin{equation}
\label{eq_aout}
    \aout_{\alpha} = \sqrt{\eta} \left( e^{i \Phi_{\alpha}} \ain_{\alpha} + ig_{\alpha} \sqrt{\kappa} \,\chi_{\alpha} \xb \right) + \sqrt{1-\eta}\, a^v_{\alpha} 
\end{equation}
where $\ain_{\alpha} $ denotes the vacuum input field, $\Phi_{\alpha}$ a frequency-dependent phase, $\chi_{\alpha}$ the cavity susceptibility, and $a^v_{\alpha}$ the vacuum field associated with the overall propagation and detection losses, quantified by the efficiency $\eta$. This expression highlights that, besides enabling cooling and stabilization, the outgoing field $A$ also acts as a meter for the bright mode displacement $\xb$ (here normalized to its zero-point fluctuations).  

The $B$ field, blue detuned with respect to the nearest cavity resonance, plays the central role in the experiment. When $\Delta_B \simeq \Wb$, its interaction with the nanosphere motion is well described by a two-mode squeezing Hamiltonian $\hbar \gB \left(\aB \hat{b} + \aB^{\dagger} \hat{b}^{\dagger} \right)$, which generates entanglement between the electromagnetic field and the bright mechanical mode. These quantum correlations are preserved in the field propagating out of the cavity. 
Inside the cavity, the electromagnetic field modes are uniquely determined by the boundary conditions. In contrast, propagating modes allow for a certain degree of arbitrariness in their definition. We may introduce a mode annihilation operator as 
\begin{equation}
\label{eq_axi}
a_{\xi} = \int_{-\infty}^{\infty} \xi(-t) \aout  (t) \ud t = \int_{-\infty}^{\infty} \xi(\omega) \aout (\omega) \frac{\ud \omega}{2\pi}
\end{equation}
where the envelope function $\xi(t)$ is normalized according to $\int_{-\infty}^{\infty} |\xi(t)|^2 \ud t=1$ \cite{Genes2008a}. 

It can be shown that $a_{\xi}$ obeys the canonical bosonic commutation relations $[a_{\xi}, a_{\xi}^{\dagger}]=1$. The corresponding mode quadratures are defined as $\,Q_{\xi} = a_{\xi}+ a^{\dagger}_{\xi}$ and $\,P_{\xi}=i (a^{\dagger}_{\xi}- a_{\xi})$.
In the following we will consider propagating optical modes defined by a flat spectral window centered at $-\Wb$ with width $\Gammaxi$.

\textit{Data analysis and modeling} - Intracavity operators are not directly accessible, but the system state is encoded in the fields exiting the cavity, and we retrieve this information using heterodyne detection. To this purpose, two local oscillators, detuned by $\WALO/2\pi=1.4\, \mathrm{MHz}$ and $\WBLO/2\pi=2.0\,\mathrm{MHz}$ are combined and directed onto a single balanced detection that probes the light transmitted by the cavity.
The heterodyne photocurrent operator is $i_{het}= \alphaLO^{*}\, \aout \,e^{i \WLO t} +\alphaLO \,a^{\scriptscriptstyle{out}\,\dagger} \,e^{-i \WLO t}$ where $\alphaLO$ is the local oscillator field~\cite{Bowen2015}. $i_{het}$ is Hermitian, hence it satisfies $i_{het} (-\omega) = (i_{het}(\omega))^{\dagger}$.  

From the heterodyne signal we experimentally reconstruct the spectral correlation matrix $\matCA=\langle \mathrm{a_i} \, \mathrm{a_j} \rangle$ of the output fields ladder operators $\mathbf{a}=\left(\aout_A ,a_A^{\scriptscriptstyle{out}\,\dagger}, \aout_B , a_B^{\scriptscriptstyle{out}\,\dagger}   \right)$ where $O^\dagger(\omega)=\mathcal{F}[O^\dagger(t)]$ \cite{SM1,Zippilli2015}. To this end, a phase-sensitive analysis is required. For this reason the nanosphere is positioned at a small but finite distance from nodes of both $A$ and $B$ intracavity stationary waves, so that weak coherent signals are generated in $\aout_{\alpha}$ providing a stable phase reference and enabling consistent determination of the field quadratures across the full data set. 

The heterodyne signal is continuously acquired and segmented into 10 ms long intervals. For each interval, the Fourier transform of the heterodyne signal, $v(\omega)$, is computed and assembled into the four-dimensional data vector
$\mathbf{v}(\Omega)= \bigl(v^{*}(\WALO-\Omega), v(\WALO+\Omega), v^{*}(\WBLO-\Omega), 
v(\WBLO+\Omega)\bigr)$.
From this vector, compensated for the frequency-dependent argument of the electronic response, we evaluate the spectral correlation matrix $\mathbf{C}$ defined as $\mathrm{C}_{ij}=\langle \mathrm{v}_i(-\Omega) \mathrm{v}_j(\Omega) \rangle$ where $\langle \cdot\rangle$ denotes ensemble averaging. Calibration of  $\mathbf{C}$ to the standard quantum noise level of fields $A$ and $B$ (obtained by blocking the signal path in the heterodyne detection) is followed by subtraction of electronic noise. This procedure transforms $\mathbf{C}$  into a new matrix $\mathbf{A}^\phi=\mathbf{R} \mathbf{A}\mathbf{R}^{T}$ where $\mathbf{R}$ is a constant matrix accounting for the phases, $\phi_A$ and $\phi_B$, of the two detected coherent components with respect to the intracavity fields, which define the reference phase in the model.

\begin{figure}[!htb]
    \includegraphics[width=\textwidth]{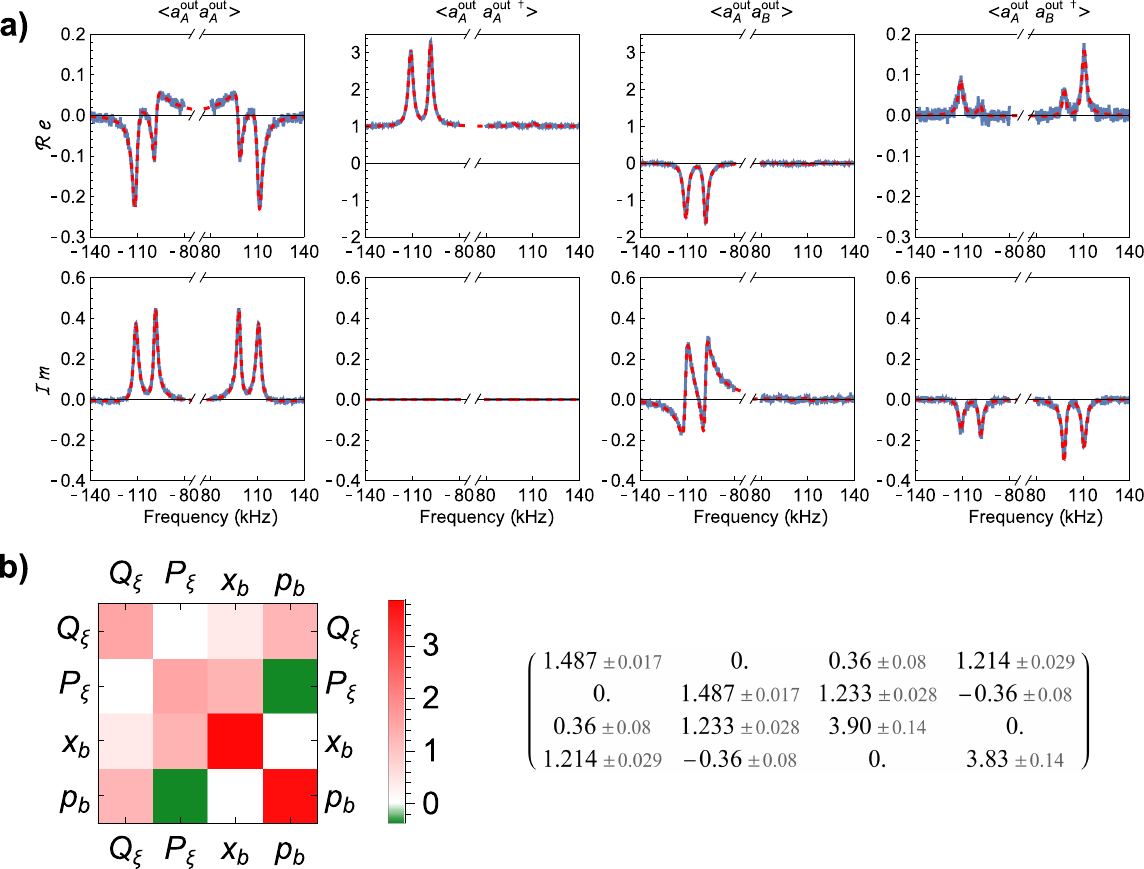}
    \caption[Fig2]{\textbf{Optomechanical correlations}. \textbf{a)} First row of the complex-valued spectral matrix $\matCA$, shown around the bright mode sidebands, along with the fitted theoretical model (dashed-red line).  Each column in the figure represent the real (top) and imaginary (bottom) part of $\matCA$.  \textbf{b)} 2D portrayal of the covariance matrix $\matV$ fully characterizing the optical-mechanical Gaussian state, and corresponding numerical values. The estimated dynamical parameters for this dataset are $\Wb/2\pi=106\pm0.1$\,kHz, $g_A/2\pi=11.7\pm0.2$\,kHz, $g_B/2\pi=6.3\pm0.1$\,kHz. The heating rates are $\Gamma_x/2\pi=3.00\pm0.05$\,kHz and $\Gamma_y/2\pi=2.67\pm0.05$\,kHz. All quoted errors indicate one SD of $\sim10$ independent samples. }
    \label{Fig2}
\end{figure}

The spectral matrix $\matCA$  contains all the information that can be extracted from the two outgoing fields. Owing to intrinsic symmetries, it comprises 16 different real-valued, frequency-dependent functions.
The elements $\mathrm{A}^\phi_{1,2}$ and $\mathrm{A}^\phi_{3,4}$ represent the usual heterodyne power spectral density (PSD) of the fields $A$ and $B$, respectively. These are fitted simultaneously to the optomechanical model to extract the system parameters, namely, the pairs of mechanical bare frequencies,  heating rates, and optomechanical coupling rates. Independent measurements provide $\kappa/2\pi=58\pm0.6\,\mathrm{kHz}$ and $\eta= 0.283\pm0.006$. The full $\mathbf{A}^\phi$ matrix is then fitted to the model to determine  $\phi_A$ and $\phi_B$. Representative examples of elements of $\matCA$ are shown in Fig.~\ref{Fig2}~a) around the bright mode sidebands, along with the respective fitting curves.

The close agreement between the experimental data and the curves produced by the optomechanical model over the entire spectral matrix indicates a precise quantitative understanding of the system.
This enables us to employ the
same model, with the fitted parameters, to compute the covariance matrix $\matV$ describing the joint state of the bright mechanical mode and the propagating field mode $B$. Its elements are defined as $\mathrm{V}_{ij} = 0.5\langle \{\mathrm{u}_i , \mathrm{u}_j\} \rangle$, where $\mathbf{u} = (Q_{\xi},\,  P_{\xi},\, \xb,\, \pb)$. Here $\xb$ and $\pb$ denote the position and momentum of the bright mode normalized to their respective zero-point fluctuations, and $\{\cdot,\cdot \}$ indicates the anti-commutator.

We return to this method for evaluating 
$\matV$ below. First, however, we consider a more direct route that fully exploits the matrix $\matCA$, which provides a complete description of the correlations between the output fields $A$ and $B$. This is achieved by applying a suitable linear transformation and integrating over frequency. The transformation matrix incorporates the envelope function $\xi(\omega)$, while the mechanical observables are reconstructed from field $A$ by inverting Eq.~\ref{eq_aout}. Explicitly, we define $\xm = \frac{\aout_{A}}{i\sqrt{\eta\,\kappa}\,g_{A}  \,\chi_{A}} $, $\ppm = -i \frac{\omega}{\Wb} \xm$, which can be used as measured estimators of the mechanical quadratures $x_b$ and $p_b$ when evaluating $\matV$ \cite{SM1}. 
Importantly, the noise properties and correlations that quantify the quantum character of the system
emerge directly from the processed data,
instead of being deduced from the fit of the model.
A representative covariance matrix obtained with this method is shown in Fig.~\ref{Fig2}\,b). 

\begin{figure}[!htb]
    \centering
      \includegraphics[scale=1.0]{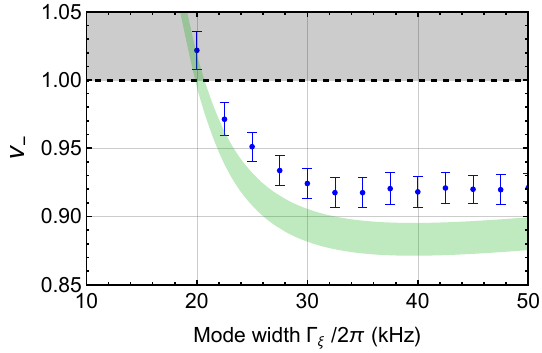}
    \caption[Fig3]{\textbf{Optimization of the optical mode width.} The smallest symplectic eigenvalue $\num$ of the partially trasposed covariance matrix $\matV$ is plotted as a function of the mode width $\Gamma_\xi$. The data points with error bars (indicating the SEM over about 10 independent data acquisitions) are obtained from $\matV$ directly reconstructed from the experimental correlation matrix. The green shaded area derives from $\matV$ computed with our optomechanical model, using the parameters that best fit the heterodyne spectra. The width of this band reflects systematic uncertainties in the model parameters (mainly the efficiency $\eta$), with negligible contribution from statistical uncertainty. The gray shaded region highlights the threshold above which the mechanical and optical subsystems become separable.}
    \label{Fig3}
\end{figure}

\textit{Results} \& \textit{discussion} - The entanglement between the optical and mechanical subsystems is quantified by the smallest symplectic eigenvalue, $\num$, of the partially transposed covariance matrix \cite{Duan2000,Simon2000}. With our normalization, $\num = 1$ if the two oscillators are in uncorrelated vacuum states, and the presence of entanglement is signaled by $\num < 1$. Fig.~\ref{Fig3} reports the values of $\num$ calculated from the 
covariance matrix $\matV$ directly reconstructed through $\matCA$,  as a function of the width $\Gammaxi$ of $\xi(\omega)$. Entanglement emerges when the bandwidth $\Gammaxi$ of the optical mode becomes sufficiently large to encompass the full spectral response of the mechanical bright mode~\cite{Genes2008a}, characterized by two peaks at $\sim\WX$ and $\sim\WY$ broadened by optomechanical interaction. The minimum value of $\num$, corresponding to maximal entanglement, is achieved around $\Gammaxi/2\pi =35\,$kHz and reaches $\num=0.918 \pm 0.029$. Here, the quoted statistical uncertainty represents one standard deviation obtained from multiple measurements, while an additional systematic uncertainty of $\pm 0.02$ arises from the independent estimates of $\eta$ and $\kappa$. 

Entanglement is commonly quantified by the logarithmic negativity $E_{N}$, defined as $E_{N} = max\{0, -\log_2(\num)\}$~\cite{Vidal2002,Plenio2005}. At its maximum, we obtain $E_{N} = 0.12\pm0.04$, including both the statistical and the systematic uncertainty.

Fig.~\ref{Fig3} also reports the values of $\num$ obtained by estimating $\matV$ from the full optomechanical model using all the fitted parameters, including the heating rates. The model reproduces the overall trend of the values extracted directly from the experimental covariance matrix, but consistently yields lower values of $\num$, with a minimum of $\num=0.884\pm0.011$ ($E_{N} = 0.177\pm0.019$).  We attribute this discrepancy to slow fluctuations of the system parameters during the acquisition of the full time series, most notably of the relative phases $\phi_A$ and $\phi_B$, which are not fully captured by the stationary description. To mitigate these effects, the data analysis is restricted to time intervals over which the fitted parameters remain approximately constant. In practice, each independent estimate of the correlation matrix  $\matCA$ — and hence of the covariance matrix  $\matV$ and the symplectic eigenvalue  $\num$ — is obtained by averaging over typically 2000 consecutive time windows of $10$\,ms. Residual parameter fluctuations during this 20 s acquisition time nevertheless mix slightly different Gaussian states and reduce the observable entanglement. As a consequence, the entanglement actually present in the system is more faithfully captured by estimating $\matV$ through the optomechanical model evaluated with the fitted parameters, whereas the values of $\num$ directly inferred from the data should be regarded as a conservative estimate of the entanglement that can be effectively harnessed under the present level of system stability. 

We also note that the rectangular spectral filter used here is not optimal. Tailored mode shaping could increase the opto-mechanical fidelity and enhance the observable entanglement~\cite{Bowen2015}.

\begin{figure}[!htb]
    \centering
      \includegraphics[scale=1.0]{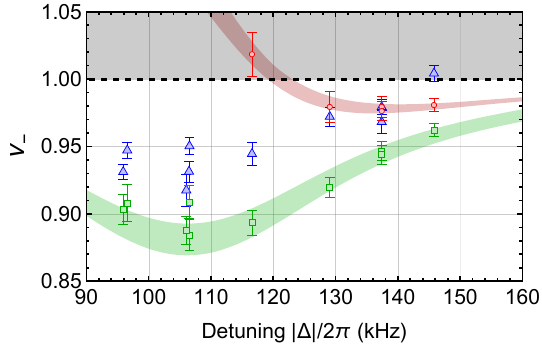}
    \caption[Fig4]{\textbf{Symplectic eigenvalue versus cavity detuning.} 
    Experimental data points shown by blue markers with error bars (SEM over about 10 independent data acquisitions) are obtained from the covariance matrix directly reconstructed from the measured correlations. Green markers are extracted instead from the same experimental data through the optomechanical model, using the parameters that best fit the heterodyne spectra. In this case the uncertainty is dominated by the systematic error on the detection efficiency $\eta$.
    In these measurements, we set $-\Delta_A\simeq\Delta_B=|\Delta|$. Intracavity entanglement is inferred from the symplectic eigenvalues associated to the covariance matrix calculated, within the same model and parameter set, for the joint state of the motion along an optimized direction and the intracavity $B$ field (red points, with error bars again set by the systematic uncertainty on $\eta$). The green and red shaded regions are obtained by taking the extremal model predictions generated by the full set of parameters extracted from all data sets (at all detunings). These bands therefore illustrate the overall detuning dependence of the entanglement in our system.}
    \label{Fig4}
\end{figure}

The results discussed so far were obtained for detunings satisfying $-\Delta_A \simeq \Delta_B \simeq \Wb$. This condition, however, is not essential, and a finite degree of entanglement is present even when it is relaxed. Indeed, we repeated the experiment while varying the detuning, keeping only the condition $-\Delta_A \simeq \Delta_B$.  Fig.~\ref{Fig4} shows the symplectic eigenvalue $\num$ 
derived from the covariance matrix $\matV$ reconstructed directly from the measured correlation matrix $\matCA$, together with the corresponding values obtained by computing $\matV$ with the optomechanical model using parameters fitted to the same experimental data. 

Entanglement persists over a detuning range exceeding 40 kHz, consistent with the expectation that it is maintained within a bandwidth on the order of the cavity linewidth $\kappa$. This robustness implies that stable opto-mechanical entanglement can be achieved without fine control of the detuning. 

An important and nontrivial question is whether entanglement is also present inside the cavity, between the intracavity mode of field $B$ and the mechanical oscillator. This is reasonable but not obvious, since the intracavity field mode is uniquely determined by the cavity boundary conditions and cannot be tailored as in the case of the propagating field.
Although intracavity observables cannot be accessed directly, the same optomechanical model that quantitatively reproduces the measured output correlations allows us to reconstruct the corresponding intracavity covariance matrix from the fitted parameters. A crucial issue is the motion along the tweezer axis (the $z$ mode), whose mechanical frequency is $\Omega_z/2\pi = 19.3\,$kHz. This motion is coupled to the cavity field due to a small angle between the cavity axis and the tweezer transverse plane. As a consequence, the $z$ mode produces a peak also in the spectrum of the bright mode through the cavity-mediated coupling, slightly increasing its variance. We have indeed added this contribution in the entries $\mathrm{V}_{33}$ and $\mathrm{V}_{44}$ before computing the values of $\num$ here reported \cite{SM1}. The $z$ motion has no effect on the propagating optical mode, because the filter $\xi$ removes
the spectral region around the $z$ mode resonance. In contrast, the $z$ mode peak is not negligible in the spectra of the cavity optical modes and, in fact, it destroys the entanglement with the transverse mechanical bright mode considered so far. Nevertheless, optomechanical entanglement is partially recovered by considering a slightly tilted direction of motion, thereby including a projection of the tweezer axis. This enhances the optomechanical correlations in the spectral region of the $z$ mode.   
The symplectic eigenvalue $\num$ associated with the covariance matrix describing the cavity $B$ mode and the optimized motion direction, shown in Fig.~\ref{Fig4} as a function of the detuning, confirms the presence of intracavity entanglement, albeit at a reduced level compared to that of the propagating mode. The minimum inferred value of $\num$ for the intracavity variables is $\num = 0.976\pm0.007$, corresponding to a logarithmic negativity of $E_N = 0.035\pm0.010$, and occurs at a detuning of $|\Delta|/2\pi = 137$\,kHz.

In conclusion, we demonstrate quantum entanglement between the center-of-mass motion of a levitated nanosphere and a propagating optical field. By combining coherent scattering in an optical cavity with high efficiency heterodyne detection, we reconstruct the full set of optical–mechanical correlations and observe the violation of separability bounds under experimentally accessible conditions. These results establish levitated optomechanical systems as a versatile platform for generating and distributing light–matter entanglement at room temperature, without ultra-cryogenic environments or stringent parameter tuning,
providing a significant contribution to macroscopic quantum technologies.
Beyond center-of-mass motion, additional mechanical degrees of freedom — including rotational and librational modes~\cite{Barker2023,kamba2023,Dania2025,Toros2025} — provide further opportunities for multimode quantum control.
Combined with the prospect of trapping and simultaneously cooling multiple levitated nanospheres in a single optical cavity \cite{Vijayan2024,Pontin2025}, this architecture could enable multipartite entanglement mediated by a shared optical field \cite{Chauhan2022}.
More broadly, the ability to prepare nonclassical states of increasingly massive objects may open the way to experimental tests of the possible quantum nature of gravity~\cite{Bose2017,Marletto2017} and conjectured modifications of quantum mechanics at macroscopic scales.

\section*{Acknowledgments}
The authors would like to acknowledge useful discussions with D. Vitali, P. F. Barker and U. Deli\'{c}. We acknowledge financial support from PNRR MUR Project No. PE0000023-NQSTI and by the European Commission-EU under the Infrastructure I-PHOQS “Integrated Infrastructure Initiative in Photonic and Quantum Sciences ” [IR0000016, ID D2B8D520, CUP D2B8D520].

\bibliography{database}

\end{document}


\setcounter{figure}{0}
\renewcommand{\thefigure}{S\arabic{figure}}

\setcounter{equation}{0}
\renewcommand{\theequation}{S\arabic{equation}}

\title[]{Stationary entanglement of a levitated oscillator with an optical field}

\author{Q. Deplano}%
\affiliation{Dipartimento di Fisica e Astronomia, Università degli Studi di Firenze, via Sansone 1, I-50019 Sesto Fiorentino, Italy}%
\affiliation{INFN, Sezione di Firenze, via Sansone 1, I-50019 Sesto Fiorentino, Italy}

\author{A. Pontin}
\email{antonio.pontin@cnr.it}
\affiliation{CNR-INO, largo Enrico Fermi 6, I-50125 Firenze, Italy}

\author{F. Marino}%
\affiliation{CNR-INO, largo Enrico Fermi 6, I-50125 Firenze, Italy}
\affiliation{INFN, Sezione di Firenze, via Sansone 1, I-50019 Sesto Fiorentino, Italy}

\author{F. Marin}
\email{francesco.marin@unifi.it}
\affiliation{Dipartimento di Fisica e Astronomia, Università degli Studi di Firenze, via Sansone 1, I-50019 Sesto Fiorentino, Italy}%
\affiliation{INFN, Sezione di Firenze, via Sansone 1, I-50019 Sesto Fiorentino, Italy}
\affiliation{CNR-INO, largo Enrico Fermi 6, I-50125 Firenze, Italy}
\affiliation{European Laboratory for Non-Linear Spectroscopy (LENS), Via Carrara 1, I-50019 Sesto Fiorentino, Italy}

\maketitle

\tableofcontents

\newpage

\section{Experimental setup}
\begin{figure}[h]
\includegraphics[width=13.5cm]{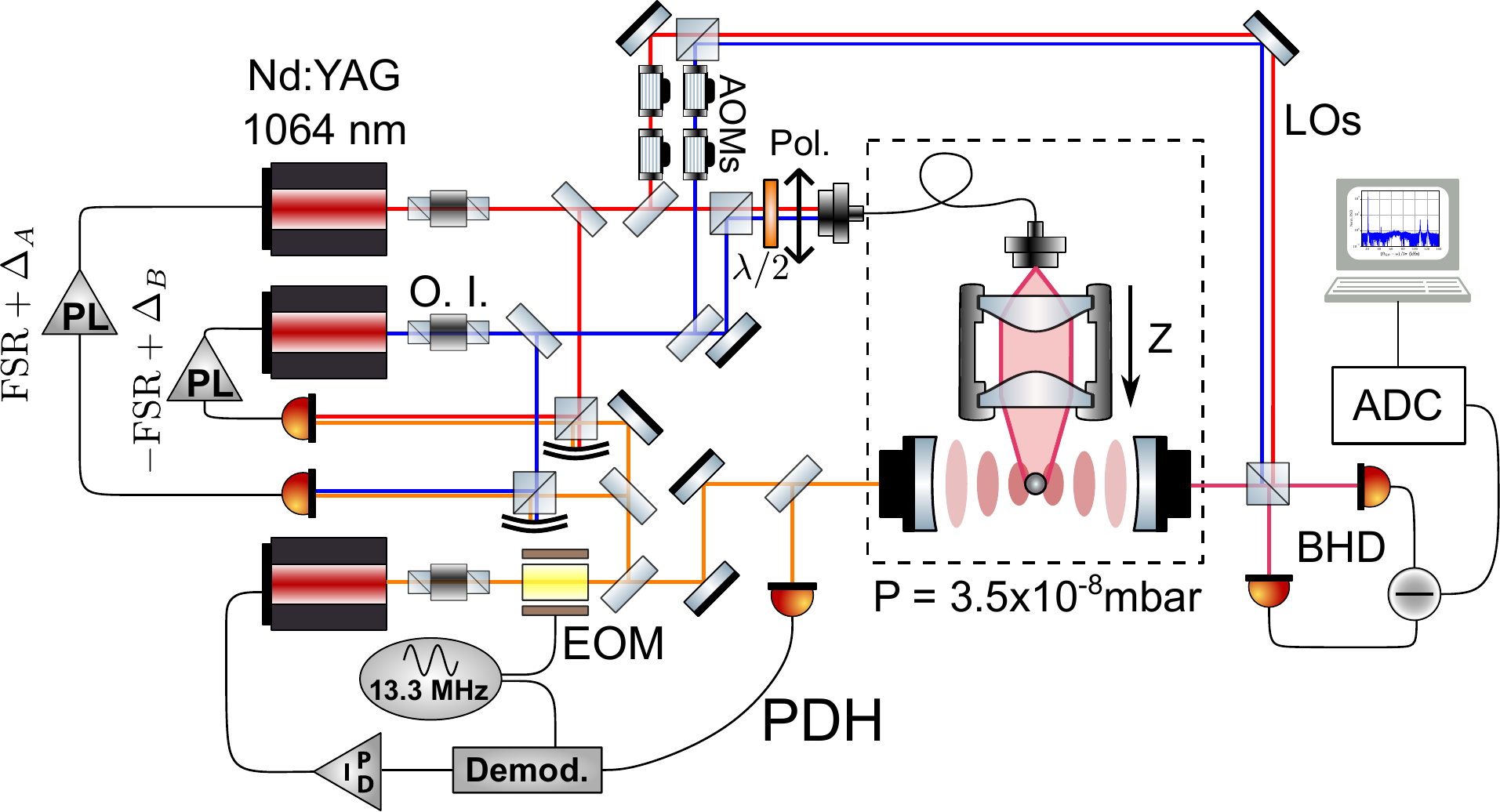}
\caption{Optical layout. O.I: optic isolator, AOM: acousto-optic modulator, EOM: resonant electro-optic modulator, FSR: free-spectral range, PDH: Pound-Drever-Hall detection, LO: local oscillator, BHD: balanced heterodyne detection, ADC: analog-to-digital converter, PL: phase/frequency offset lock.
}\label{fig1sup}
\end{figure}

The optical layout of the experiment is shown in Fig.~\ref{fig1sup}.  A first Nd:YAG laser at $1064$\,nm is locked to the optical cavity by implementing a Pound-Drever-Hall (PDH) scheme and serves as a frequency reference. The light from two additional Nd:YAG lasers, denoted $A$ and $B$ and used to form the optical tweezer, is mixed with two fields obtained by splitting the reference beam, each combination being detected with a fast photodiode. The resulting beat notes are processed by two independent phase/frequency offset locking circuits, acting on the respective laser sources. The offset frequencies are chosen such that each laser is slightly detuned with respect to two different cavity resonances, adjacent to the reference one. The detunings $\Delta_A$ and $\Delta_B$ can be finely adjusted. For the purpose of this work, laser $B$ is blue-detuned from the cavity resonance, creating a two-mode squeezing interaction that generates quantum correlations between the particle center-of-mass motion and the optical mode. Laser $A$ is red-detuned, and used to cool the center-of-mass motion, stabilize the nanosphere, and measure its position. Portions of the $A$ and $B$ beams are frequency shifted by means of acousto-optic modulators, overlapped, and sent as local oscillators (LO) to a balanced heterodyne detection (BHD) analyzing the tweezer light scattered by the nanosphere and transmitted through the cavity.  

The description of the nanosphere loading procedure can be found in Ref. \cite{Calamai2021}. Further details on the optical cavity and its characterization, the tweezer optical setup, the detection system and its calibration can be found in the Supplemental Informations of Ref. \cite{Ranfagni2021}.

\section{Model}
\subsection{Cavity optomechanical system}
\label{Sec_IIa}

We consider the  cavity optomechanical system composed of the two optical modes ($A$ and $B$) and three mechanical modes describing the motion of the nanosphere in the tweezer transverse plane, with axes $X$ and $Y$, and along the tweezer axis $Z$. The motion is described in terms of dimensionless position and momentum operators $\xX, \pX, \xY, \pY, \xZ, \pZ$ obtained by normalizing the corresponding physical variables to their respective zero-point fluctuations $\xzpf = \sqrt{\hbar/2 m \Omegai}$ and $\pzpf = \sqrt{\hbar m \Omegai/2 }$ ($m$ is the mass of the nanosphere, $\Omegai$ the oscillation frequencies of the $i$ mode in the absence of optomechanical interaction, with $i = x, y, z$). The fields are described by the quadratures $\,Q_{\scriptscriptstyle{\alpha}}\,=\,(a_{\scriptscriptstyle{\alpha}}\,+\,a_{\scriptscriptstyle{\alpha}}^{\dag})\,$ and  $\,P_{\scriptscriptstyle{\alpha}}\,=\,i(a_{\scriptscriptstyle{\alpha}}^{\dag}\,-\,a_{\scriptscriptstyle{\alpha}})\,$ with $\alpha = (A, \, B)$. The Langevin equations are: 
\begin{eqnarray} 
\dot{x}_{\scriptstyle{i}} & = & \Omegai \,\ppi 
\label{eq_dxdt}\\[8pt]
\dot{p}_{\scriptstyle{i}} &=& -\Omegai \,\xxi + 2 \sum_{\alpha} \galphai\, \Qa + 2 \sqrt{\Gi} \,\zetai \\[6pt]
\dot{Q}_{\scriptscriptstyle{\alpha}} &=& -\Delta_{\scriptscriptstyle{\alpha}} \Pa -\frac{\kappa}{2} \,\Qa + \sqrt{\kappa}\, Q^{\scriptscriptstyle{in}}_{\scriptscriptstyle{\alpha}}  \\[6pt]
\dot{P}_{\scriptscriptstyle{\alpha}} &=& \Delta_{\scriptscriptstyle{\alpha}} \Qa -\frac{\kappa}{2}\, \Pa + 2 \sum_{i} \galphai \,\xxi + \sqrt{\kappa}\, P^{\scriptscriptstyle{in}}_{\scriptscriptstyle{\alpha}} 
\label{eq_dPdt}
\end{eqnarray}
where $\,Q_{\scriptscriptstyle{\alpha}}^{\mathrm{in}}\,=\,(a_{\scriptscriptstyle{\alpha}}^{\mathrm{in}}\,+\,a_{\scriptscriptstyle{\alpha}}^{\mathrm{in}\, \dagger})\,$ and  $\,P_{\scriptscriptstyle{\alpha}}^{\mathrm{in}}\,=\,i(a_{\scriptscriptstyle{\alpha}}^{\mathrm{in}\, \dagger}\,-\,a_{\scriptscriptstyle{\alpha}}^{\mathrm{in}})\,$ are the quadratures of the input vacuum noise, $\Delta_{\scriptscriptstyle{\alpha}}$ is the detuning from the cavity mode (i.e., for each field $\Delta = 2 \pi (\nu_{\scriptscriptstyle{\mathrm{L}}}-\nu_{\scriptscriptstyle{\mathrm{c}}})$ where $\nu_{\scriptscriptstyle{\mathrm{L}}}$ is the laser field frequency and $\nu_{\scriptscriptstyle{\mathrm{c}}}$ the cavity mode resonance frequency), $\kappa$ is the cavity width. The noise sources have the only non-null correlation functions $\langle \zetai(t) \, \zetai(t')\rangle =  \langle a_{\scriptscriptstyle{\alpha}}^{\mathrm{in}}(t) \, a_{\scriptscriptstyle{\alpha}}^{\mathrm{in}\,\dagger}(t')\rangle= \delta(t-t')$. Due to the high background temperature, we are classically modeling the mechanical noise. 
The gas damping plays a negligible role at low pressure, where optomechanical damping largely dominates, and we neglect it in the model. The decoherence of the motion is mainly due to collisions with the background gas, and to quantum noise in the dipole scattered light. The corresponding heating rates $\Gi$ are left as free parameters in the analysis of the experimental data. The optomechanical coupling rates $\galphaX$ and $\galphaY$ are proportional to $\sin^2 \theta_p$ and $\sin \theta_p \cos \theta_p$ respectively, where $\theta_p$ is the angle between the cavity axis and the tweezer polarization axis. The two laser fields used for the tweezer have the same polarization at the input of the tweezer optical fiber, defined by a polarizer after they have been overlapped. As a consequence, the ratio $\galphaX/\galphaY$ is the same for the two fields. We simplify the model by defining $\gX = \gAX$, $\gY = \gAY$ (both free parameters, determined by the fits to the data), and assuming $\gBX = \sqrt{\Prat} \gX$, $\gBY = \sqrt{\Prat} \gY$ where $\Prat$ is a further free parameter, accounting for the ratio between the  
$B$ and $A$ laser powers in the tweezer.

The quantum steady state of the optomechanical system is characterized by its covariance matrix $\mathbf{V}^{\mathrm{c}}$ with elements $\mathrm{V}^{\mathrm{c}}_{ij} = 0.5\langle \{\mathrm{u}_i , \mathrm{u}_j\} \rangle$ where $\mathbf{u}^{\mathrm{T}} = (\QA,\,  \PA,\, \QB,\,  \PB,\, \xX,\, \pX,\, \xY,\, \pY,\,\xZ,\,\pZ )$.  It can be calculated using the Lyapunov equation $\,\Drift\, \mathbf{V}^{\mathrm{c}}+\mathbf{V}^{\mathrm{c}} \,\Drift^{\mathrm{T}} = -\mathbf{D}$ where the drift matrix $\Drift$ and the diffusion matrix $\mathbf{D}$, derived from the Langevin model, are
\begin{equation}
    \Drift=
    \left(
    \begin{array}{cccccccccc}
        -\kappa/2 & -\Delta_{\scriptscriptstyle{A}} & 0  & 0 & 0  & 0 & 0 & 0& 0 & 0\\
         \Delta_{\scriptscriptstyle{A}} & -\kappa/2 & 0 & 0 & 2 \gX  & 0 & 2 \gY & 0 & 2 \gAZ & 0\\
         0  & 0 &-\kappa/2 & -\Delta_{\scriptscriptstyle{B}} &  0  & 0 & 0 & 0& 0 & 0\\
         0  & 0 &\Delta_{\scriptscriptstyle{B}} & -\kappa/2 & 2 \sqrt{\Prat}\gX  & 0 & 2 \sqrt{\Prat}\gY & 0 & 2 \gBZ & 0 \\
         0 & 0 & 0 & 0 & 0  & \OmegaX & 0  & 0 & 0 & 0\\
         2 \gX & 0 & 2 \sqrt{\Prat}\gX & 0 & -\OmegaX  & 0 & 0  & 0 & 0 & 0\\
         0 & 0 & 0 & 0 & 0  & 0 & 0  & \OmegaY & 0 & 0\\
         2 \gY & 0 & 2 \sqrt{\Prat}\gY & 0 & 0  & 0 & -\OmegaY  & 0 & 0 & 0\\
          0 & 0 & 0 & 0 & 0  & 0 & 0  & 0 & 0 & \WZ\\
         2 \gAZ & 0 & 2 \gBZ & 0 & 0  & 0 & 0  & 0 & -\WZ & 0\\
    \end{array}
    \right)
\end{equation}
and $\mathbf{D} = \mathrm{Diag} [\kappa,\,\kappa,\,\kappa,\,\kappa,\,0,\,4\GX,\,0,\,4\GY\,0,\,4\GZ]$.

It is convenient to define a rotated frame in the transverse plane of the tweezer, with one axis determined by  the projection on this plane of the cavity axis. Hence, we define the bright (b) and dark (d) modes as 
\begin{eqnarray}
    x_\mathrm{b} &=& \left(\sqrt{\Wb/\WX}\,\sin{\theta}\right) \xX + \left(\sqrt{\Wb/\WY}\,\cos{\theta}\right) \xY \\
    x_\mathrm{d} &=& \left(\sqrt{\Wd/\WX}\,\cos{\theta}\right) \xX - \left(\sqrt{\Wd/\WY}\,\sin{\theta}\right) \xY
\end{eqnarray}
    where 
    \begin{eqnarray}
    \Wb &=& \sqrt{\WX^2\, \sin^2\theta + \WY^2\, \cos^2 \theta}  \\ 
    \Wd &=& \sqrt{\WX^2\, \cos^2\theta + \WY^2\, \sin^2 \theta} 
    \end{eqnarray}
    and the angle $\theta$ is formally defined as $\tan\theta = \sqrt{\frac{\WX}{\WY}}\frac{\gX}{\gY}$. If the coupling rates $\gX$ and $\gY$ depend on the polarization angle as described above, then $\theta \equiv \theta_p$ and $x_\mathrm{b}$ represents the motion along the projection of the cavity axis. We notice that $x_\mathrm{b} = \frac{1}{g_\mathrm{b}}\left(\gX x+ \gY y\right)$, where the optomechanical coupling rate for the bright mode is defined as 
    \begin{equation}
    \gb = \sqrt{(\gX^2 \WX + \gY^2 \WY)/\Wb}    \, .
    \end{equation}
    The transformation matrix describing the rotation of the reference frame, including the correct normalization, is

\begin{equation*}
    \Rbd=
    \left(
    \begin{array}{cccccccccc}
        1 & 0 & 0  & 0 & 0  & 0 & 0 & 0 & 0 & 0\\
        0 & 1 & 0 & 0 & 0  & 0 &0 & 0 & 0 & 0\\
         0  & 0 &1 & 0 &  0  & 0 & 0 & 0 & 0 & 0\\
         0  & 0 & 0 & 1  & 0 & 0 & 0 & 0 & 0 & 0\\
         0 & 0 & 0 & 0 & \sqrt{\Wb/\WX}\,\sin{\theta}  & 0 & \sqrt{\Wb/\WY}\,\cos{\theta}  & 0 & 0 & 0\\
         0 & 0 & 0  & 0 & 0  & \sqrt{\WX/\Wb}\,\sin{\theta} & 0 & \sqrt{\WY/\Wb}\,\cos{\theta} & 0 & 0\\
         0 & 0 & 0  & 0 & \sqrt{\Wd/\WX}\,\cos{\theta}  & 0 & -\sqrt{\Wd/\WY}\,\sin{\theta} & 0 & 0 & 0\\
         0 & 0 & 0  & 0 & 0  & \sqrt{\WX/\Wd}\,\cos{\theta} & 0 & -\sqrt{\WY/\Wd}\,\sin{\theta} & 0 & 0\\
          0 & 0 & 0  & 0 & 0  & 0 & 0 & 0 & 1 & 0\\
        0 & 0 & 0 & 0 & 0  & 0 &0 & 0  & 0 & 1\\
    \end{array}
    \right) \quad .
\end{equation*}

For the purpose of maximizing the intracavity entanglement between the $B$ cavity mode and a mode of the three-dimensional motion, we will also consider a further rotation of the coordinate frame around the axis of the dark mode, i.e., in the plane defined by the cavity axis and the $Z$ axis. This rotation is parametrized by an angle $\thz$ and defines two mechanical modes, $\bphi$ and $\zphi$, with  eigenfrequencies
    \begin{eqnarray}
    \Wzb &=& \sqrt{\Wb^2\, \sin^2\thz + \WZ^2\, \cos^2 \thz}  \\ 
    \Wbz &=& \sqrt{\Wb^2\, \cos^2\thz + \WZ^2\, \sin^2 \thz} \,.
    \end{eqnarray}
The corresponding transformation matrix is given by

\begin{equation*}
    \Rbz=
    \left(
    \begin{array}{cccccccccc}
        1 & 0 & 0  & 0 & 0  & 0 & 0 & 0 & 0 & 0\\
        0 & 1 & 0 & 0 & 0  & 0 &0 & 0 & 0 & 0\\
         0  & 0 &1 & 0 &  0  & 0 & 0 & 0 & 0 & 0\\
         0  & 0 & 0 & 1  & 0 & 0 & 0 & 0 & 0 & 0\\
         0 & 0 & 0 & 0 & \sqrt{\Wbz/\Wb}\,\cos{\thz}  & 0 & 0 & 0 & \sqrt{\Wbz/\WZ}\,\sin{\thz}   & 0\\
         0 & 0 & 0  & 0 & 0  & \sqrt{\Wb/\Wbz}\,\cos{\thz} & 0 & 0 & 0& \sqrt{\WZ/\Wbz}\,\sin{\thz} \\  
          0 & 0 & 0  & 0 & 0  & 0 & 1 & 0 & 0 & 0\\
        0 & 0 & 0 & 0 & 0  & 0 &0 & 1  & 0 & 0\\
                0 & 0 & 0  & 0 & -\sqrt{\Wzb/\Wb}\,\sin{\thz}  & 0 & 0 & 0  & \sqrt{\Wzb/\WZ}\,\cos{\thz} & 0 \\
         0 & 0 & 0 & 0  & 0  & -\sqrt{\Wb/\Wzb}\,\sin{\thz} & 0 & 0 & 0 & \sqrt{\WZ/\Wzb}\,\cos{\thz} \\
    \end{array}
    \right) \quad .
\end{equation*}
If the optomechanical coupling of the motion along $Z$ originates entirely from its projection on the cavity axis, then a rotation $\thz$ equal to $\thz_0 =\arctan{\sqrt{\frac{\WZ}{\Wb}}\frac{\gZ}{\gb}}$ aligns the mode $\bphi$ with the cavity axis.

The covariance matrix in the rotated mechanical frame is 
\begin{equation}\label{eq_covvlyap}
\mathbf{V}^{\mathrm{bd}} = \Rbz\,\Rbd\, \mathbf{V}^{\mathrm{c}}\, \Rbd^{\mathrm{T}} \, \Rbz^{\mathrm{T}}\,.
\end{equation}
Finally, the optomechanical subsystem composed of the $B$ cavity field and the $\bphi$ mechanical mode, that exhibits entanglement, is described by selecting the third to sixth rows and columns of $\mathbf{V}^{\mathrm{bd}}$, i.e., by the covariance matrix $\mathbf{V}^{\mathrm{Bb}}$ whose elements are $\mathrm{V}^{\mathrm{Bb}}_{ij} = \mathrm{V}^{\mathrm{bd}}_{i+2,j+2}\,$ with $1 \leqslant i,j \leqslant 4$. An example of the matrix $\mathbf{V}^{\mathrm{bd}}$ is reported in Section \ref{Sec_intracavity}.

\subsection{Propagating fields and spectral correlation matrix}
\label{Sec_propagating}

To better emphasize the physical effects most relevant to our system, we now restrict our analysis to the two-dimensional motion in the transverse plane of the tweezer. The longitudinal $z$ mode will be incorporated at a later stage, in Section \ref{Sec_axial}.  The Langevin equations governing the intracavity dynamics can be written in the Fourier space as
\begin{eqnarray}
\aA(\omega) &=& \chiA (\omega) \left[i \gb \,\xb(\omega)+\sqrt{\kappa} \,\aAin (\omega)  \right]
\label{eq_Lang1} \\
\adA(\omega) &=& \chiA^* (-\omega) \left[-i \gb \,\xb(\omega)+\sqrt{\kappa} \,\adAin (\omega)  \right]
\label{eq_Lang2} \\
\aB(\omega) &=& \chiB (\omega) \left[i \sqrt{\Prat}\gb \,\xb(\omega)+\sqrt{\kappa} \,\aBin (\omega)  \right]
\label{eq_Lang3} \\
\adB(\omega) &=& \chiB^* (-\omega) \left[-i \sqrt{\Prat}\gb \,\xb(\omega)+\sqrt{\kappa} \,\adBin (\omega)  \right]
\label{eq_Lang4} \\
\xb(\omega) &=& \frac{2}{\gb}  \left( \gX^2 \,\chi_x + \gY^2 \,\chi_y \right) \left(\aA(\omega)+\adA(\omega) +\sqrt{\Prat} \left(\aB(\omega) +\adB (\omega)\right)  
+ N(\omega)  \right)
\label{eq_Lang5}  \\
\pb\om &=& -i\frac{\omega}{\Wb}\,\xb(\omega)
\label{eq_Lang6}
\end{eqnarray}
where Fourier transformed ($\mathcal{F}$) operators are defined as $O\left(\omega\right)\equiv\mathcal{F}\left[O \left(t\right)\right]=\int O(t) e^{i\omega t} \ud t$, and
$O^{\dagger}\left(\omega\right)=\mathcal{F}\left[O^{\dagger}\left(t\right)\right]$. In the following, we will use power spectral densities (PSD) of the operators, defined as $S_{OO}=\int \langle O^{\dagger}(\omega')\,O(\omega)\rangle\frac{\ud \omega'}{2\pi}$, and cross spectral densities defined as $S_{O_1 O_2}=\int \langle O_1^{\dagger}(\omega')\,O_2(\omega)\rangle\frac{\ud \omega'}{2\pi}$. 
We have also defined the susceptibilities
\begin{eqnarray}
\chi_j\left(\omega\right) &=& \frac{\Omega_j}{\Omega_j^2-\omega^2}\\
\label{eq_chialpha}
\chialpha\left(\omega\right) &=& \frac{1}{-i
\left(\Delta_{\scriptscriptstyle{\alpha}}+\omega\right)+\kappa/2}
\end{eqnarray}
with $j= (x, y)$, and written the mechanical noise input in the compact form
\begin{gather}
N(\omega) = \frac{\gX \sqrt{\GX}\, \chi_x \,\zeta_x\om\,+\,\gY \sqrt{\GY}\, \chi_y \,\zeta_y\om}{\gX^2 \,\chi_x + \gY^2 \,\chi_y }   \, .
\end{gather}
The field $\aoc$ at the output of the cavity is given by the input-output relation 
\begin{equation}
\label{eq_aout}
\aoc = \sqrt{\kappa} a -\ain, 
\end{equation}
and the overall losses can be included in the propagating (and detected) output field by introducing an additional vacuum field $a^{\scriptscriptstyle{v}}$ and an efficiency $\eta$, so that $\aout_{\alpha} = \sqrt{\eta}\,\aoc_{\alpha} + \sqrt{1-\eta}\, a^{\scriptscriptstyle{v}}_{\alpha} $. Due to the overlap between the optical paths of the $A$ and $B$ fields and their close wavelengths, we have assumed that the efficiency $\eta$ is the same for the two fields. Using Eq. (\ref{eq_Lang1},\ref{eq_Lang3}) we thus obtain Eq. (1) of the main text, i.e.,
\begin{equation}
\label{eq_det}
    \aout_{\alpha} = \sqrt{\eta} \left( e^{i \Phi_{\alpha}} \ain_{\alpha} + ig_{\alpha} \sqrt{\kappa} \,\chi_{\alpha} x_b \right) + \sqrt{1-\eta}\, a^{\scriptscriptstyle{v}}_{\alpha} 
\end{equation}
where we have expressed in a compact form the cavity induced dephasing as $e^{i \Phi_{\alpha}} = \kappa\, \chi_{\alpha} -1$, and we have defined $\gA = \gb,\,\gB =\sqrt{\Prat} \gb$ .

Eqs. (\ref{eq_Lang1}-\ref{eq_Lang6}), together with Eq. (\ref{eq_aout}) and its Hermitian conjugate, establish a set of linear equations for the variables $\mathbf{a}^{\scriptscriptstyle{oc}} = \left( \aAoc, \,\adAoc, \,\aBoc, \,\adBoc, \,\xb, \,\pb \right)^{\mathrm{T}} $ with input noise terms $\mathbf{a}^{\scriptscriptstyle{in}} = \left( \aAin, \,\adAin, \,\aBin, \,\adBin, \zeta_x,\, \zeta_y\right)^{\mathrm{T}}$, that is readily solvable in the form $\mathbf{a}^{\scriptscriptstyle{oc}} = \mathbf{M}\, \mathbf{a}^{\scriptscriptstyle{in}}$ where $\mathbf{M}(\omega)$ is a deterministic matrix. In particular, the expression for $\xb(\omega)$, which yields the fifth row of $\mathbf{M}$ is
\begin{equation}
    \label{eq_xb}
    \xb = \chieff(\omega)\, 2 \gb \left[\sqrt{\kappa} \left(\chiA(\omega) \aAin+\chiA^{*}(-\omega)\adAin +\sqrt{\Prat}\, \chiB(\omega)\aBin +\sqrt{\Prat}\,\chiB^{*}(-\omega)\adBin   \right)
    + N(\omega)\right]
\end{equation}
where we have defined an effective susceptibility 
\begin{equation}
    \label{eq_chieff}
    \chieff(\omega) = \frac{\gX^2\,\chi_x +\gY^2 \,\chi_y  }{\gb^2 \left[1-2 \,i\, \left( \gX^2\, \chi_x + \gY^2 \,\chi_y \right)\bigl( \chiA(\omega)-\chiA^{*}\left(-\omega\right)+\Prat\,\chiB(\omega)-\Prat\,\chiB^{*}\left(-\omega\right)\bigr)  \right]}  
\end{equation}
and the remaining entries of $\mathbf{M}(\omega)$ follow directly by substituting $\xb$ into Eqs. (\ref{eq_Lang1}-\ref{eq_Lang4},\ref{eq_Lang6},\ref{eq_aout}).

The input noise terms have non-null correlation functions $\langle \aalphain(\omega')\,\adalphain(\omega)\rangle=\langle \zeta_i(\omega')\,\zeta_i(\omega)\rangle =2\pi \,\delta(\omega'+\omega)$. 
The spectral correlation matrix for the output variables, defined as $\mathrm{A}^{\scriptscriptstyle{oc}}_{ij}(\omega) = \int\langle \mathrm{a}^{\scriptscriptstyle{oc}}_i(\omega')\,\mathrm{a}^{\scriptscriptstyle{oc}}_j (\omega)\rangle\,\frac{\ud\omega'}{ 2\pi}$, has therefore the form
\begin{equation}
\label{eq_Aoc}
\mathbf{A}^{\scriptscriptstyle{oc}}(\omega) = \mathbf{M}(-\omega) \mathbf{\Cout} \mathbf{M}^{\mathrm{T}}(\omega) 
\end{equation}
with
\begin{equation*}
    \mathbf{\Cout}=
    \left(
    \begin{array}{cccccc}
        0 & 1 & 0  & 0 & 0  & 0\\
        0 & 0 & 0 & 0 & 0  & 0 \\
         0  & 0 &0 & 1 &  0  & 0 \\
         0  & 0 & 0 & 0  & 0  & 0\\
  0  & 0 & 0 & 0  & 1  & 0\\
  0  & 0 & 0 & 0  & 0  & 1
    \end{array}
    \right) \quad .
\end{equation*}

If we now consider the detected field, including losses, instead of the cavity output fields, the vector of the output variables is $\mathbf{a}^{\scriptscriptstyle{out}} = \left( \aAo, \,\adAo, \,\aBo, \,\adBo, \,\xb, \,\pb \right)^{\mathrm{T}} $ and the correlation matrix takes the form 
\begin{equation}
    \label{eq_Aout}
\mathbf{A}^{\scriptscriptstyle{out}}= \Deta\,\mathbf{A}^{\scriptscriptstyle{oc}}\,\Deta + (1-\eta)\,\mathbf{\Cout}\,\mathbf{I}_{\mathrm{o}}
\end{equation}
where $\Deta = \mathrm{Diag} [\sqrt{\eta},\,\sqrt{\eta},\,\sqrt{\eta},\,\sqrt{\eta},\,1,\,1]$ 
and $\mathbf{I}_{\mathrm{o}} = \mathrm{Diag} [1,\,1,\,1,\,1,\,0,\,0]$. The correlation matrix that can be directly measured, called $\mathbf{A}$ in the main text, corresponds to the first four rows and columns of $\mathbf{A}^{\scriptscriptstyle{out}}$.  

\subsection{Detection}\label{sup_detection}

In a heterodyne detection of a the output field $\aout$, the photocurrent operator is $i_{het}(t)= \alphaLO^{*}\, \aout(t) \,e^{i \WLO t} +\alphaLO \,a^{\scriptscriptstyle{out}\,\dagger}(t) \,e^{-i \WLO t}$ where $\alphaLO$ is the local oscillator field and $\WLO$ is its detuning from the signal field (i.e., $\WLO = 2 \pi (\nu_{\scriptscriptstyle{\mathrm{LO}}}-\nu_{\scriptscriptstyle{\mathrm{L}}})$ where $\nu_{\scriptscriptstyle{\mathrm{LO}}}$ is the frequency of the local oscillator). In the Fourier space, 
$i_{het}(\omega) =\alphaLO^{*}\,\aout(\omega+\WLO)+\alphaLO \,\adout(\omega-\WLO)$. In the following, we assume that the photocurrent is normalized to the shot-noise level, and that the phase of the local oscillator with respect to the intracavity field — which is the reference frame in this model — is determined via a global fit of the correlation matrix. Hence, we set $\alphaLO=1$. The operator $i_{het}(t)$ is Hermitian, therefore it satisfies $i_{het} (-\omega) = (i_{het}(\omega))^{\dagger}$.  In particular,
\begin{eqnarray}
\label{eq_ihet1}
    i_{het}(\WLO+\Omega)  &=& \adout(\Omega) \,+\,\aout(\Omega+2\WLO)  \\
    \label{eq_ihet2}
    \left[i_{het}(\WLO+\Omega)\right]^{\dagger}  &=&\aout(-\Omega) \,+\,\adout(-\Omega-2\WLO)  \, .
\end{eqnarray}
Since $\WLO \gg \kappa, \Wb$, the right-end terms in Eqs. (\ref{eq_ihet1}-\ref{eq_ihet2}) correspond to vacuum states, therefore we can write the spectral correlations as
\begin{eqnarray}
    \label{eq_ihetcorr1}
    \langle i_{het}(\WLO+\Omega')\,i_{het}(\WLO+\Omega)\rangle &=&\langle \adout(\Omega')\,\adout(\Omega)\rangle \\
    \label{eq_ihetcorr2}
    \langle i_{het}(\WLO+\Omega')\,[i_{het}(\WLO-\Omega)]^{\dagger}\rangle &=&\langle \adout(\Omega')\,\aout(\Omega)\rangle +\delta(\Omega'+\Omega)\\
    \label{eq_ihetcorr3}
    \langle [i_{het}(\WLO-\Omega')]^{\dagger}\,i_{het}(\WLO+\Omega)\rangle &=&\langle \aout(\Omega')\,\adout(\Omega)\rangle \\
    \label{eq_ihetcorr4}
    \langle [i_{het}(\WLO-\Omega')]^{\dagger}\,[i_{het}(\WLO-\Omega)]^{\dagger}\rangle &=&\langle \aout(\Omega')\,\aout(\Omega)\rangle
\end{eqnarray}
If we now consider our double heterodyne detection, performed with the same balanced detection on the output fields $A$ and $B$, Eqs. (\ref{eq_ihetcorr1}-\ref{eq_ihetcorr4})  hold for both $A$ and $B$, but an additional shot-noise term $\,\delta(\Omega+\Omega')\,$ must be added on the right-end side of Eqs. (\ref{eq_ihetcorr2}, \ref{eq_ihetcorr3}). We call $v$ the total photocurrent of the balanced detection, define the vector 
$\,\mathbf{v}(\Omega)= \bigl( v^{*}(\WALO-\Omega),\,v(\WALO+\Omega),\, v^{*}(\WBLO-\Omega),\,
v(\WBLO+\Omega)\bigr)^{\mathrm{T}}\,$ (where $v^{*}(\omega)$ is the complex conjugate of $v(\omega)$), and consider the spectral correlation matrix $\mathbf{C}$ with elements 
\begin{equation}
    \label{eq_C}
\mathrm{C}_{ij}=\langle \mathrm{v}_i(-\Omega) \mathrm{v}_j(\Omega) \rangle \,  
\end{equation}
where $\langle \cdot\rangle$ denotes ensemble averaging. By integrating Eqs. (\ref{eq_ihetcorr1}-\ref{eq_ihetcorr4}) over $\Omega'$ and considering the delta-correlated noise sources, we obtain that $\mathrm{C}_{ij} = \mathrm{A}_{ij} + 2 (\delta_{21}+\delta_{43})+\delta_{12}+\delta_{34}$, where $\delta_{ij}$ is the Kronecker delta.

The elements $\mathrm{A}_{1,2}(\Omega)\equiv S_{\mathrm{het}}^{\scriptscriptstyle{A}}(\WALO+\Omega)$ and $\mathrm{A}_{3,4}(\Omega)\equiv S_{\mathrm{het}}^{\scriptscriptstyle{B}}(\WBLO+\Omega)$ represent the usual heterodyne power spectral density of the fields $A$ and $B$, respectively. Their explicit expressions are 
\begin{eqnarray}
S_{\mathrm{het}}^{\scriptscriptstyle{A}}(\WALO-\Omega) &=& 1+\eta \,\kappa\, |\chiA(\Omega)|^2\,\gb^2 \,S_{\xb \xb}(\Omega)  
\label{eq_ShetA}
\\ S_{\mathrm{het}}^{\scriptscriptstyle{B}}(\WBLO-\Omega) &=& 1+\eta \,\kappa\, |\chiB(\Omega)|^2\,\gb^2 \,\Prat\,S_{\xb \xb}(\Omega)
\label{eq_ShetB}
\end{eqnarray}
where the power spectrum of $\xb$ is 
\begin{equation}
\label{eq_Sxb}
    S_{\xb \xb}(\omega) = 4 \gb^2 \,|\chieff(\omega)|^2 \,\left[ \kappa\, |\chiA(-\omega)|^2 + \kappa\,\Prat\, |\chiB(-\omega)|^2  +\frac{\gX^2\,\GX\,|\chi_x|^2 + \gY^2\,\GY\,|\chi_y|^2 }{|\gX^2\,\chi_x + \gY^2 \, \chi_y |^2}
    \right]  \, .
\end{equation}

The experimental $\mathrm{A}_{1,2}$ 
and $\mathrm{A}_{3,4}$ are fitted simultaneously to the optomechanical model to extract the system parameters (see Sec.~\ref{supp_par_est}). 

\subsection{Propagating field mode and covariance matrix}

As described in the main text, we introduce a propagating optical mode with annihilation operator  
\begin{equation}
\label{eq_axi}
a_{\xi} = \int_{-\infty}^{\infty} \xi(-t) \aout  (t) \ud t = \int_{-\infty}^{\infty} \xi(\omega) \aout (\omega) \frac{\ud \omega}{2\pi}
\end{equation}
where the envelope function $\xi(t)$ is normalized according to $\int_{-\infty}^{\infty} |\xi(t)|^2 \ud t=1$. The corresponding mode quadratures are defined as 
\begin{eqnarray}
Q_{\xi} &=& a_{\xi}+ a^{\dagger}_{\xi} = \int_{-\infty}^{\infty} \bigl( 
\xi(\omega) \aout (\omega)
+ \xi^{*}(-\omega) \adout (\omega)
\bigr) \frac{\ud \omega}{2\pi}\\
P_{\xi} &=& i \,(a^{\dagger}_{\xi}- a_{\xi})  = i \int_{-\infty}^{\infty} \bigl( 
-\xi(\omega) \aout (\omega)
+ \xi^{*}(-\omega) \adout (\omega)
\bigr) \frac{\ud \omega}{2\pi}  \, .
\end{eqnarray}
We can define the vector of the spectral quadratures of the mechanical and optical modes as $\,\mathbf{q}^{\scriptscriptstyle{out}}(\omega) = \mathbf{M}^{\scriptscriptstyle{\xi}} (\omega)\,\mathbf{a}^{\scriptscriptstyle{out}} (\omega)\,$ where
\begin{equation*}
    \mathbf{M}^{\scriptscriptstyle{\xi}} (\omega)=\left(
    \begin{array}{cccccc}
        \xi(\omega) & \xi^{*}(-\omega) & 0  & 0 & 0  & 0\\
        -i\,\xi(\omega) & i\,\xi^{*}(-\omega) & 0 & 0 & 0 & 0  \\
         0  & 0 &\xi(\omega) & \xi^{*}(-\omega) &  0  & 0 \\
         0  & 0 & i\,\xi(\omega) & -i\,\xi^{*}(-\omega)  & 0  & 0\\
  0  & 0 & 0 & 0  & 1  & 0\\
  0  & 0 & 0 & 0  & 0  & 1
    \end{array}
    \right) \quad .
\end{equation*}
We are using the same envelope function of the $A$ and $B$ modes to simplify the notation, but we are only interested in the $B$ mode. The covariance matrix $\mathbf{V}^{\scriptscriptstyle{\xi}}$ for the variables $\mathbf{q}^{\scriptscriptstyle{\xi}} = (Q_{\xi,\scriptscriptstyle{A}},\,P_{\xi,\scriptscriptstyle{A}},\,Q_{\xi,\scriptscriptstyle{B}},\,P_{\xi,\scriptscriptstyle{B}},\,\xb,\,\pb)^{\mathrm{T}}\,$ is defined as
\begin{equation}
    \mathrm{V}^{\scriptscriptstyle{\xi}}_{ij} = \frac{1}{2} \langle \mathrm{q}^{\scriptscriptstyle{\xi}}_i\,\mathrm{q}^{\scriptscriptstyle{\xi}}_j+\mathrm{q}^{\scriptscriptstyle{\xi}}_j\,\mathrm{q}^{\scriptscriptstyle{\xi}}_i \rangle = \frac{1}{2}\int \frac{\ud \omega'}{2\pi} \int \frac{\ud \omega}{2\pi}\,\langle \mathrm{q}^{\scriptscriptstyle{out}}_i(\omega')\,\mathrm{q}^{\scriptscriptstyle{out}}_j(\omega)+\mathrm{q}^{\scriptscriptstyle{out}}_j(\omega')\,\mathrm{q}^{\scriptscriptstyle{out}}_i(\omega) \rangle \, .
\end{equation}
It can be written using the matrix $\mathbf{S}$ of the cross spectral densities of $\mathbf{q}^{\scriptscriptstyle{\xi}}$, whose elements are defined as $\,\mathrm{S}_{ij}(\omega) =\int \frac{\ud \omega'}{2\pi}\langle \mathrm{q}^{\scriptscriptstyle{out}}_i(\omega')\,\mathrm{q}^{\scriptscriptstyle{out}}_j(\omega) \rangle$,  in the form
\begin{equation}
\label{eq_Veta_a}
     \mathbf{V}^{\scriptscriptstyle{\xi}} = \frac{1}{2}\int \frac{\ud \omega}{2\pi} \,\left( \mathbf{S}(\omega) + \mathbf{S}^{\mathrm{T}} (\omega) \right) \, .
\end{equation}
$\mathbf{S}$ in turn can be expressed as a function of the correlation matrix $\mathbf{A^{\scriptscriptstyle{out}}}$, according to $\mathbf{S}(\omega) = \mathbf{M}^{\scriptscriptstyle{\xi}}(-\omega)\, \mathbf{A}^{\scriptscriptstyle{out}}(\omega)\,\mathbf{M}^{\scriptscriptstyle{\xi}\,\scriptstyle{\mathrm{T}}}(\omega) $ 
or, using Eqs. (\ref{eq_Aoc}, \ref{eq_Aout}), 
\begin{equation}
\label{eq_Veta_c}
   \mathbf{S}(\omega) =\Deta\,\mathbf{M}^{\scriptscriptstyle{\xi}}(-\omega) \mathbf{M}(-\omega)\mathbf{\Cout}  \mathbf{M}^{\scriptstyle{\mathrm{T}}}(\omega) \mathbf{M}^{\scriptscriptstyle{\xi}\,\scriptstyle{\mathrm{T}}}(\omega)\, \Deta + (1-\eta)\, \mathbf{I}_{\mathrm{o}}\,|\xi(-\omega)|^2  \, .
\end{equation}

Once the parameters of the optomechanical system have been determined by fitting the heterodyne spectra, the matrix 
$\mathbf{M}$ can be constructed. Using Eqs. (\ref{eq_Veta_a},\ref{eq_Veta_c}), the corresponding covariance matrix is then obtained. For our purposes, the relevant portion is the block $\mathbf{V}^{\scriptscriptstyle{\xi B}}$ defined by the last four rows and columns of $\mathbf{V}^{\scriptscriptstyle{\xi}}$, which describes the optomechanical subsystem consisting of the propagating $B$ mode and the mechanical bright mode.

An example of the matrix $\mathbf{V}^{\scriptscriptstyle{\xi B}}$ is given by

\begin{equation}
    \label{eq:eq_example_VB}
    \mathbf{V}^{\scriptscriptstyle{\xi B}}=
    \begin{pmatrix}
        1.5 & 0. & 0.35 & 1.29 \\
        0. & 1.5 & 1.31 & -0.36 \\
        0.35 & 1.31 & 3.89 & 0. \\
        1.29 & -0.36 & 0. & 3.81
    \end{pmatrix}
\end{equation}

\noindent where the following parameters have been considered  $(\OmegaX,\OmegaY,\gX,\gY,\GX,\GY,\Delta_A,\Delta_B)/2\pi=(110.5,\, 98.3,\, 9.0,\, 7.5,\, 3.0,\, 2.7,\, -98.6,\, 96.6)$\,kHz and
$\Prat=0.291$. They correspond to one of the experimentally determined parameters set.

\subsection{Mechanical variables measured via the $A$ field}
\label{Sec_mechanical}

A different approach is considering the field $A$ as a meter for the mechanical bright mode, and derive the covariance matrix from the measured spectral correlation matrix $\mathbf{A}$. To this purpose, we define $\,\xm = \frac{\aout_{A}}{i\sqrt{\eta\,\kappa}\,\gb \,\chi_{A}} \,$ and $\,\ppm = -i \frac{\omega}{\Wb} \xm$. With respect to $\xb$ and $\pb$, they include additional terms proportional to $\aAin$ and $a^{\scriptscriptstyle{v}}_{A}$, whose effect must be evaluated. 

We first consider the mechanical subsystem, starting from $\langle \xm\,\xm \rangle$. The only non-null additional contribution is 
\begin{eqnarray*}
    \langle \xm(\omega')\,\xm(\omega) - \xb(\omega')\,\xb(\omega) \rangle &=& \frac                 {i(\Delta_{\scriptscriptstyle{A}}+\omega')+\kappa/2}{i\,\gb\,\sqrt{\kappa}}\,\langle \aAin(\omega')\,\xb(\omega)\rangle \\
     &=& -2 i  \chieff(\omega)\frac{i(\Delta_{\scriptscriptstyle{A}}+\omega')+\kappa/2}{i(\Delta_{\scriptscriptstyle{A}}-\omega)+\kappa/2} \langle \aAin (\omega')\,\adAin (\omega) \rangle
\end{eqnarray*}
where in the second step we have used the expression (\ref{eq_xb}) for $\xb$ and kept only the non-null terms of the correlation. By integrating over $\omega'$ we obtain
\begin{equation}
    S_{\xm \xm}-S_{\xb \xb}=
    -2 \,i\, \chieff(\omega)=-2\,i\,|\chieff(\omega)|^2 \,\frac{1}{\chieff(-\omega)}
\end{equation}
where in the last step we have used $\chieff(\omega)=\chieff^{*}(-\omega)$. While the spectrum of $\xb$ is real, that of $\xm$ is not, due to the additional non-Hermitian term. However, we can consider its real part, and using the expression of $\chialpha$ and $\chieff$ given in Eq. (\ref{eq_chialpha}, \ref{eq_chieff}) we obtain an additional contribution $S_{\scriptscriptstyle{\mathrm{add}}} = \mathrm{Re} \left[S_{\xm \xm}\right]-S_{\xb \xb}$ in the form
\begin{equation}
\label{eq_Sadd}
S_{\scriptscriptstyle{\mathrm{add}}} = 4 \gb^2\, |\chieff(\omega)|^2\,\kappa\, \frac{1}{2}\bigl( |\chiA(\omega)|^2+\Prat\,|\chiB(\omega)|^2-|\chiA(-\omega)|^2-\Prat\,|\chiB(-\omega)|^2\bigr) \, .
\end{equation}
Comparing $S_{\scriptscriptstyle{\mathrm{add}}}$ with $ S_{\xb \xb}$ (see Eq. (\ref{eq_Sxb})) we find that the additional term symmetrizes the spectrum of $S_{\xb \xb}$, while leaving unchanged its frequency integral that appears on the diagonal of the covariance matrix. Specifically, $\mathrm{Re} \left[ S_{\xm \xm}  \right]= S_{\xb \xb}+S_{\scriptscriptstyle{\mathrm{add}}} = 0.5 \bigl(S_{\xb \xb}(\omega)+S_{\xb \xb}(-\omega)\bigr)$. An analogous result holds for the momentum spectrum, whereas the position–momentum cross correlations remain null. 

We now consider the correlations between the propagating $B$ mode and the mechanical subsystem, starting from $\xm$ and $Q_{\xi}$. The only non-null additional contribution with respect to $1/2 \,\left(S_{\xb Q_{\xi}}+S_{Q_{\xi} \xb}\right)$ is
\begin{eqnarray*}
    S_{13}^{\scriptscriptstyle{\mathrm{add}}} &=&\int \frac{\ud \omega'}{2\pi}\,\frac{-1}{\gb \sqrt{\kappa}}\left(\kappa-\frac{1}{\chiA(\omega')}\right) \langle \aAin(\omega')\,\bigl(\xi(\omega) \aBo (\omega) + \xi^{*} (-\omega) \adBo (\omega) \bigr) \rangle \\
    &=& \sqrt{\frac{\eta \Prat}{\kappa}}\int \frac{\ud \omega'}{2\pi}\,\left(\kappa-\frac{1}{\chiA(\omega')}\right) \bigl( \xi(\omega)\chiB(\omega)  -\xi^{*}(-\omega)\chiB^{*}(-\omega) \bigr) \langle \aAin(\omega') \xb(\omega)\rangle \\
    &=& 2 \sqrt{\eta}\, \gb \sqrt{\Prat}\,\bigl( \xi(\omega)\chiB(\omega)  -\xi^{*}(\omega)\chiB^{*}(\omega) \bigr) 
\end{eqnarray*}
where in the second step we have used Eq. (\ref{eq_det}) for $\aBo$ and $\adBo$, and taken into account that the $B$ output field is correlated to $\aAin$ only via $\xb$. In the third step, we have replaced to $\xb$ its term proportional to $\adAin$, i.e., $\,2 \gb \sqrt{\kappa}\,\chieff(\omega)\,\chiA^{*}(-\omega)\,\adAin(\omega)\,$ (see Eq. (\ref{eq_xb})), and integrated considering the delta-correlated $A$ input field. The last expression of $S_{13}^{\scriptscriptstyle{\mathrm{add}}}$ has a symmetric imaginary part and an anti-symmetric real part with respect to $\omega$. Hence, by taking $\mathrm{Re}\left[S_{13}^{\scriptscriptstyle{\mathrm{add}}}\right]$ we have a null contribution to the covariance matrix.

The same consideration holds for the correlation between $\xm$ and $P_{\xi}$, which gives an additional contribution $S_{14}^{\scriptscriptstyle{\mathrm{add}}}\,\propto\,i\,\bigl( \xi(\omega)\chiB(\omega)  +\xi^{*}(\omega)\chiB^{*}(\omega) \bigr)$, as well as for the correlations involving $\ppm$, with  $S_{23}^{\scriptscriptstyle{\mathrm{add}}}\,\propto\,i\omega\,\bigl( \xi(\omega)\chiB(\omega)  -\xi^{*}(\omega)\chiB^{*}(\omega) \bigr)$ and $S_{24}^{\scriptscriptstyle{\mathrm{add}}}\,\propto\,\omega\,\bigl( \xi(\omega)\chiB(\omega)  +\xi^{*}(\omega)\chiB^{*}(\omega) \bigr)$. 

We now define the procedure to obtain the cross spectral densities matrix $\mathbf{S}^{\scriptscriptstyle{\mathrm{m}}}$ for the variables $(Q_{\xi,\scriptscriptstyle{B}},\,P_{\xi,\scriptscriptstyle{B}},\,\xm,\,\ppm)$, starting from the correlation matrix  $\mathbf{A}$. The relevant spectral quadratures are defined as $\,\mathbf{q}^{\scriptscriptstyle{\mathrm{m}}} = \matMm\,\bigl( \aAo,\,\adAo,\,\aBo,\,\adBo \bigr)^{\mathrm{T}}\,$ with
\begin{equation*}
    \matMm (\omega)=\left(
    \begin{array}{cccccc}
       0  & 0 & \xi(\omega) & \xi^{*}(-\omega)  \\
        0  & 0 & -i\,\xi(\omega) & i\,\xi^{*}(-\omega)  \\
         \frac{1}{i\sqrt{\eta\,\kappa}\,\gb  \,\chi_{A}(\omega)}  & 0 & 0  & 0   \\
         \frac{- \omega}{ \Wb\sqrt{\eta\,\kappa}\,\gb  \,\chi_{A}(\omega)}  & 0 &  0  & 0   
    \end{array}
    \right) \quad .
\end{equation*}
The cross spectral densities matrix $\mathbf{S}^{\scriptscriptstyle{\mathrm{m}}}$ is related to $\mathbf{A}$ by
\begin{equation}   
\label{eq_SBb}
\mathbf{S}^{\scriptscriptstyle{\mathrm{m}}}(\omega) = \matMm(-\omega)\,\mathbf{A}(\omega) \,\mathbf{M}^{\scriptscriptstyle{\mathrm{m}}\,\scriptstyle{\mathrm{T}}}(\omega)  \, .
\end{equation}
Finally, the covariance matrix $\mathbf{V}^{\scriptscriptstyle{\xi B}}$ is given by  
\begin{equation}
\label{eq_VBb}
\mathbf{V}^{\scriptscriptstyle{\xi B}} = \frac{1}{2}\int \frac{\ud \omega}{2\pi} \,\mathrm{Re} \left[ \mathbf{S}^{\scriptscriptstyle{\mathrm{m}}}(\omega) + \mathbf{S}^{\scriptscriptstyle{\mathrm{m}} \scriptstyle{\mathrm{T}}} (\omega)  \right]  \, .
\end{equation}
From the point of view of the model, the two expressions of $\mathbf{V}^{\scriptscriptstyle{\xi B}}$ given respectively in Eqs. (\ref{eq_Veta_a}, \ref{eq_Veta_c}) and in  Eqs. (\ref{eq_SBb}, \ref{eq_VBb}), are equivalent. However, they correspond to different experimental procedures: for the former, the full set of system parameters, including the diffusion coefficients (heating rates), are determined from the fit of the heterodyne spectra. For the latter, the matrix $\mathbf{A}$ is reconstructed experimentally, and only the parameters appearing in $\matMm$ (i.e., the entries of the deterministic drift matrix) are necessary.

\subsection{Measure of entanglement}

Entanglement in a bipartite Gaussian state is witnessed by the smallest symplectic eigenvalue $\num$ of the partially transposed covariance matrix \cite{Duan2000,Simon2000}. With our normalization, $\num = 1$ if the two oscillators are in uncorrelated vacuum states, and the presence of entanglement is signaled by $\num < 1$. 

We consider the covariance matrix  $\mathbf{V}$ of the system composed of the mechanical bright mode and the propagating optical mode $B$, and write it using 2x2 blocks {\boldmath $\alpha$}, {\boldmath $\beta$} and {\boldmath $\gamma$} in the form
\begin{displaymath}
    \mathbf{V} = 
    \left( \begin{array}{cc}
    \mbox{\boldmath $\alpha$} & \mbox{\boldmath $\gamma$} \\
    \mbox{\boldmath $\gamma$}^{\mathrm{T}}  &  \mbox{\boldmath $\beta$}
    \end{array}  \right) .
\end{displaymath}
Four symplectic invariants are defined as $I_1=\mathrm{Det}[\mbox{\boldmath $\alpha$}]$, $I_2=\mathrm{Det}[\mbox{\boldmath $\beta$}]$, $I_3=\mathrm{Det}[\mbox{\boldmath $\gamma$}]$, and $I_4=\mathrm{Det}[\mbox{\boldmath $V$}]$. With our definitions, a physical covariance matrix requires $\,I_1, I_2 \ge 1\,$ and $\,d_{\pm} \ge 1$, where the symplectic eigenvalues are defined by $\,d_{\pm} = \sqrt{(D \pm \sqrt{D^2-4 I_4} ) /2}\,$ 
with $D = I_1+I_2+2 I_3$. This condition is indeed verified by our experimentally determined matrix. 
The smallest symplectic eigenvalue of the partially transposed matrix is given by
\begin{equation}
    \label{eq_num}
    \num = \sqrt{\frac{1}{2}\left(\bar{D}-\sqrt{\bar{D}^2-4 I_4}\right)}
\end{equation}
where now $\bar{D}=I_1+I_2-2 I_3$.

\subsection{Axial mechanical mode}
\label{Sec_axial}

The nanosphere motion along the tweezer axis ($z$ mode) is weakly coupled to the cavity light due a small angle between the cavity axis and the tweezer transverse plane. This coupling is useful for weakly cooling the $z$ oscillation and thus maintaining a confined and stable system. However, the motion in the transverse plane is consequently coupled to the $z$ motion via the cavity field, and heated. This effect is not considered in the model described in Sections \ref{Sec_propagating}-\ref{Sec_mechanical}, and it is also neglected in the evaluation of the mechanical block of the covariance matrix. We examine it in this Section.

The spectrum of the $z$ mode is localized in a spectral region well distinct from that of the bright mode, and it can be written by slightly modifying Eq. (\ref{eq_xb}) (e.g., with $x \to z$, $\gY=0$, $\gb \to \gZ$). However, in the condition of very weak coupling, which is here verified, we can neglect the quantum backaction and use the Lorentzian approximation. The spectrum  is thus simplified to the form
\begin{equation}
\label{eq_Szz}
    S_{z z}(\omega) = \frac{\GZ}{(\GeffZ/2)^2 +(\omega - \WZ)^2}+\frac{\GZ}{(\GeffZ/2)^2 +(\omega + \WZ)^2}
\end{equation}
where the effective width $\GeffZ$ is determined by
\begin{equation}
    \GeffZ = 2\,\gAZ^2 \,\mathrm{Re} \left[ \chiA(\WZ)-\chiA(-\WZ) \right]\,+\,
    2\,\gBZ^2 \,\mathrm{Re} \left[ \chiB(\WZ)-\chiB(-\WZ) \right]  \, .
\end{equation}
The coupling of the $z$ mode to the intracavity field can be expressed by an additional term on the right hand side of Eq. (\ref{eq_Lang1}) of the form $i \chiA(\omega) \gAZ z(\omega)$, and similarly in Eqs. (\ref{eq_Lang2}-\ref{eq_Lang4}). Once replaced these additional terms in the Eq. (\ref{eq_Lang6}) for the bright mode, we obtain the additional contribution
\begin{equation}
\xb^{\scriptscriptstyle{\mathrm{add}}}=2 i \gb \chieff \bigl(\gAZ \,\chiA(\omega)-\gAZ\,\chiA^{*}(-\omega)+\gBZ \,\sqrt{\Prat}\,\chiB(\omega)-\gBZ\,\,\sqrt{\Prat}\chiB^{*}(-\omega)\bigr)\,z  \, .
\end{equation}
By taking the power spectrum of $\xb^{\scriptscriptstyle{\mathrm{add}}}$ and integrating over the frequency, we obtain an additional contribution to the variance of $\xb$ of the form
\begin{equation}
\label{eq_sigmaadd}
\langle \xb\xb \rangle ^{\scriptscriptstyle{\mathrm{add}}} = 8 \gb^2 \,|\chieff(\WZ)|^2\,|\bigl( \chiA(\WZ)-\chiA^{*}(-\WZ)\bigr)\,\gAZ + \bigl( \chiB(\WZ)-\chiB^{*}(-\WZ)\bigr)\,\sqrt{\Prat}\,\gBZ|^2\, \GZ/\GeffZ  \, . 
\end{equation}
The additional contribution to the variance of $\pb$ is $\langle \pb\pb \rangle ^{\scriptscriptstyle{\mathrm{add}}} =(\WZ/\Wb)^2\, \langle \xb\xb \rangle ^{\scriptscriptstyle{\mathrm{add}}}  $. 

The parameters of the $z$ mode are deduced from the heterodyne spectra. The relative peaks at $\Omega=\pm\WZ$ can be written as
\begin{equation}
\label{eq_ShetZ}
S_{\mathrm{het}}^{\scriptscriptstyle{z}}(\WalphaLO+\Omega) =1+ \eta \,\kappa\, \galphaZ^2 \,\GZ \left( \frac{|\chialpha(\WZ)|^2}{(\omega-\WZ)^2+(\GeffZ/2)^2} + \frac{|\chialpha(-\WZ)|^2}{(\omega+\WZ)^2+(\GeffZ/2)^2}\right) \, .
\end{equation}
The fit of the two heterodyne spectra to Eq. (\ref{eq_ShetZ}) allows to extract $\WZ$, $\GZ$, $\gAZ$ and $\gBZ$. This fit is discussed in Section \ref{Sec_Zexp}.

As discussed in Section \ref{Sec_entanglement}, we added the contribution of the $z$ mode to $\langle \xb\xb \rangle$ and $\langle \pb\pb \rangle$, to the appropriate diagonal entries of $\mathbf{V}^{\scriptscriptstyle{\xi B}}$ before using it to calculate $\num$. 

For what concerns the evaluation of the intracavity entanglement discussed in Section \ref{Sec_intracavity}, we employed the full three-dimensional description of the nanoparticle motion, detailed in Section \ref{Sec_IIa}, including the parameters of the $z$ mode.

\subsection{Excess laser noise}

In the model we consider quantum-limited laser noise. The effect of excess noise is indeed limited, when operating near a node of the cavity standing wave, by the weakness of the intracavity mean field \cite{Delic2020}. The validity of this assumption is confirmed by the agreement between the theoretical and the experimental spectral shapes on both detections, in the spectral region around the bright mode eigenfrequency. However, outside this region some small structures emerge from the model curves in the heterodyne spectra, which can be attributed to excess frequency noise. In this Section we analyze their effect on the variance of the bright mode.  

Excess laser noise can be modeled by input terms $(a^{n},\,a^{n \dagger})$ replacing $(\ain,\,\adin)$ in Eqs. (\ref{eq_Lang1}-\ref{eq_Lang4}), and consequently in Eq. (\ref{eq_xb}). 
This last equation gives additional contributions to the fluctuations of the bright mode in the form $2 \gb \chieff \sqrt{\kappa} \chialpha a^{n}$. In the frequency regions that we are considering, the effective susceptibility can be approximated as $\chieff \simeq \Wb/(\Wb^2-\omega^2)$. Since $4 \gb^2 << \kappa \Wb$, we can neglect the term proportional to $\xb$ in the equations for the fields, which simplify to $\aalpha = \chialpha(\omega) \sqrt{\kappa}\, a^{n}$. Since the excess noise is introduced inside the cavity, the output field acquires a contribution $\aalphaout = \sqrt{\kappa \eta}\, \aalpha$, that is measured in the heterodyne spectra as an additional spectral density $S_{n}$. The excess noise in the spectrum of the bright mode can now be written as
\begin{eqnarray}
    S^{n}_{\xb \xb} &=& \frac{4 \gb^2}{\kappa \eta} \frac{\Wb^2}{\left(\Wb^2-\omega^2 \right)^2}\,S_{n} \\
    S^{n}_{\pb \pb} &=& \frac{4 \gb^2}{\kappa \eta} \frac{\omega^2}{\left(\Wb^2-\omega^2 \right)^2}\,S_{n}
\end{eqnarray}
and their integration provides the additional contributions to the variances of $\xb$ and $\pb$.

We have measured in the heterodyne spectra some extra noise around 10 kHz, in correspondence of the servo bumps of the phase and frequency locking loops, and around 170 kHz, where the frequency noise of the $B$ laser weakly emerges from the the shot noise background. In both case, the calculated excess contribution to the variance of $\xb$ and $\pb$ is below $10^{-3}$, i.e., much less than their uncertainties, and we have neglected it.

\newpage
\section{Data analysis - $" \matCA "$ matrix determination}\label{sec_Amatrix}

\subsection{Data acquisition and calibration}
\begin{figure}[h]
\includegraphics[width=12cm]{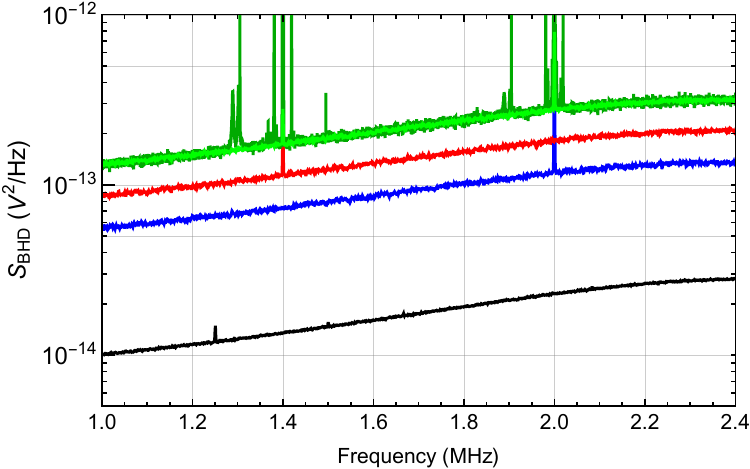}
\caption{Raw PSDs for the double heterodyne (dark-green, $10$\,s measurement time (MT)), shot noise level of LO$_A$ (red, $5$\,s MT), LO$_B$ (blue, $5$\,s MT) and dark noise level (black, $43$\,s MT). Also shown is the sum of the two shot noise levels minus the dark noise (light-green), which constitutes the heterodyne noise floor. 
}\label{fig_raw_supp}
\end{figure}

The heterodyne signal from the balanced detection (BHD in Fig.~\ref{fig1sup}) is acquired through an analog-to-digital converter ($96$\,ns sampling interval) in $2.5$\,s long sections. Acquisition of the heterodyne time traces is interlaced with the acquisition of the two local oscillators shot noise, measured independently by blocking the signal and other LO paths, as well as the detection dark (electronic) noise. The LO shot noise measurements are acquired every four heterodyne traces, while the dark noise is measured every twelve signal traces. The resulting heterodyne acquisition duty cycle is $63\%$, each shot noise measurement contributes for $16\%$ and a final $5\%$ is assigned to dark noise detection. This ensures that every $10$\,s interval of heterodyne signal acquisition has a preceding and a following shot noise measurement for each of the two LO fields. An example of the raw power spectral densities (PSDs) corresponding to a $10$\,s interval is shown in Fig.~\ref{fig_raw_supp} over a frequency band which extends across the two beat note frequencies at  $|\WALO|/2\pi=1.4\, \mathrm{MHz}$ and $|\WBLO|/2\pi=2.0\,\mathrm{MHz}$. In the same figure we show the dark noise spectrum and the shot noise levels of the two LOs averaged over a period of $5$\,s, half of which before and half after the heterodyne time traces acquisition. This strategy allows to minimize potential calibration issues due to small and slow drifts of the LOs beam power. 

In the heterodyne spectrum shown in Fig.~\ref{fig_raw_supp}, peaks due to the particle motion are clearly identifiable, at frequencies $(\Omega_x,\Omega_y,\Omega_z)/2\pi\simeq(118,98,20)$\,kHz. A calibration tone is also visible at $95$\,kHz. Data are taken with cavity detunings set at $\Delta_B \simeq -\Delta_A \simeq 2\pi\,100$\,kHz. This condition intuitively implies that the enhanced motional sidebands lie on opposite sides of the corresponding beat notes for the two fields $A$ and $B$. However, this is entirely determined by the optical frequencies of the LO beams, set by AOMs (see Fig.~\ref{fig1sup}) relative to the respective signals. In the current implementation, we have $\WALO>0$ and $\WBLO<0$ resulting in a frequency inversion in the heterodyne signal of field $B$ relative to that of field $A$.

As stated in the main text, to reconstruct the correlation matrix $\matCA=\langle \mathrm{a_i} \, \mathrm{a_j} \rangle$ it is necessary to implement a phase sensitive detection. This is done in post-processing using a numerical lock-in amplifier. The heterodyne time traces are mixed with two numerical LOs, $\propto\,\cos(\WalphaLO t)$ and $\propto\,\sin(\WalphaLO t)$ with $\alpha=(A, B)$. The results are low pass filtered (digital a-causal double low pass with cut-off at $\simeq170$\,Hz) to obtain the slowly varying quadratures of the beat notes, $X_\alpha(t)$ and $Y_\alpha(t)$. These quadratures allow us to reconstruct time traces of their instantaneous phase as $\phi_{\text{LO}} (t)=\arctan(Y(t)/X(t))$. Knowledge of $\phi_{\text{LO}}$ allows its compensation, thus achieving a fixed and stable phase reference up to an arbitrary constant $\phi_\alpha$ that is determined at a later stage. At the end of this step, one obtains two versions of the original BHD signal, $v_A$ and $v_B$, which have phase-stabilized beat notes at the respective $\WalphaLO$.

\begin{figure}[h!]
\includegraphics[width=\textwidth]{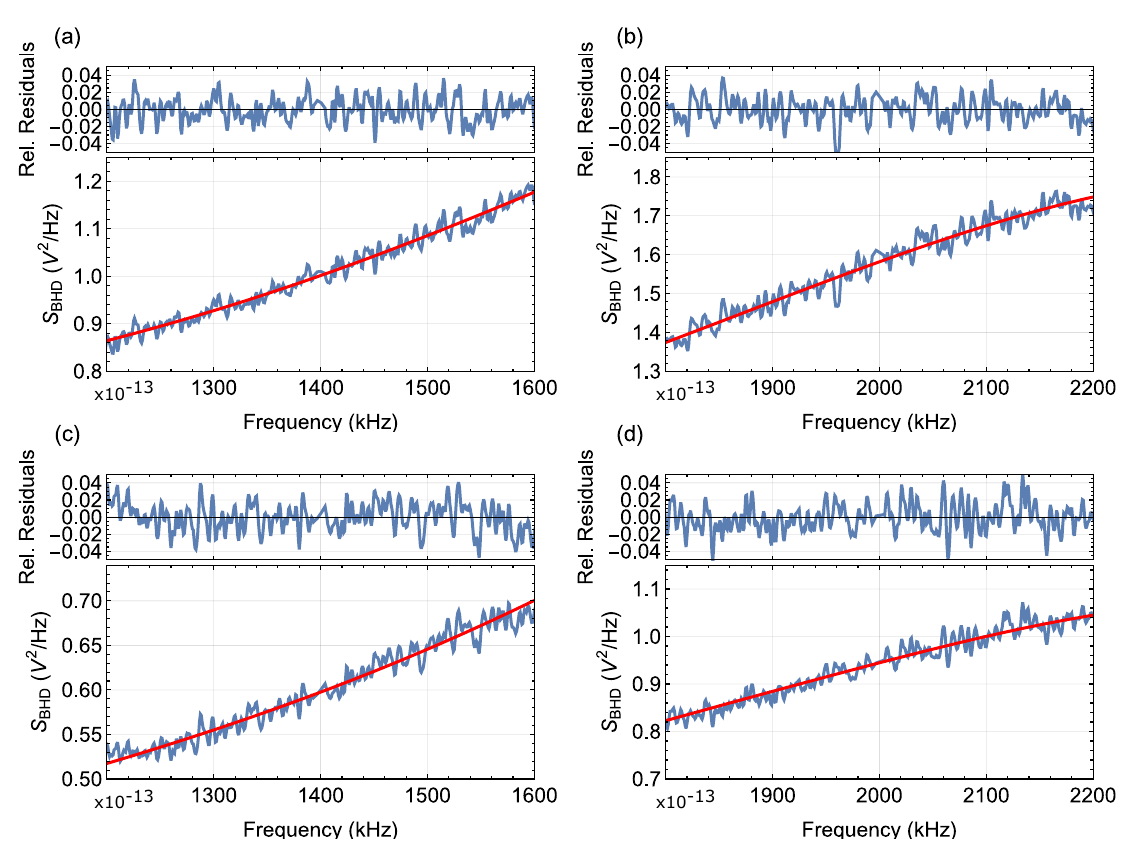}
\caption{All panels show shot noise spectra, after dark noise subtraction, and relative fit with a polynomial function. Also shown are the relative residuals. (a), (b): shot noise of $\text{LO}_\text{A}$ fitted in a bandwidth of $400$\,kHz around the two beat note frequencies. (c), (d): same as the previous panels for $\text{LO}_\text{B}$.
}\label{fig_shot_sup}
\end{figure}

The phase stabilized heterodyne signals are then segmented into $10$\,ms long intervals whose Fourier transforms, $v_A(\omega)$ and $v_B(\omega)$, are computed and re-casted in the vector $\,\mathbf{v}(\Omega)$ following the definition of the main text and Sec.~\ref{sup_detection}. It is convenient to add a frequency flip on the heterodyne of field B in order to have a consistent frequency axis defined for the two fields. Thus, in practice, we define 
$\,\mathbf{v}(\Omega)= \bigl( v_A^{*}(\WALO-\Omega),\,v_A(\WALO+\Omega),\, v_B(|\WBLO|+\Omega),\, v_B^{*}(|\WBLO|-\Omega)\bigr)\,$.  At this point, $\mathbf{v}(\Omega)$ must be corrected for the transfer function of the photodiodes (and their amplification chain) in the BHD. The argument of the transfer function has been measured independently, while the amplitude is obtained by fitting, with a polynomial function, the shot noise spectra of the two LOs in a frequency bandwidth of $400$\,kHz around the two beat notes. This implies four fits, an example of which is shown in Fig.~\ref{fig_shot_sup}. This procedure is repeated for each $10$\,s long block of heterodyne traces and simultaneously allows whitening $\mathbf{v}(\Omega)$ and calibrating it to the standard quantum noise level. We remark that we have previously shown a common mode rejection of $40$\,dB for the implemented BHD~\cite{Ranfagni2021}. This results in a residual amplitude noise at the output of the balanced detection smaller than $3\%$ of the shot noise level with the LO powers currently used. Thus, direct fitting represents a reliable measurement of the vacuum level. After whitening and calibration the ensemble average of the spectral correlation matrix $\mathbf{C}$ is calculated, following the definition of Eq. (\ref{eq_C}).

The last step to obtain the matrix $\mathbf{A}^\phi$ is the subtraction of the electronic dark noise $S_{d}(\omega)$ and the additional shot noise terms from the relevant elements of the matrix $\mathbf{C}$ (see Sec.~\ref{sup_detection}), according to

\begin{equation*}
    \mathbf{A}^{\phi}=\mathbf{C}-
    \left(
    \begin{array}{cccc}
        0  &     1+S_d(\WALO+\Omega) &    0  &     0 \\
        2+S_d(\WALO-\Omega)  &     0 &    0  &     0 \\
        0  &     0 &    0  &     1+S_d(\WBLO-\Omega) \\
        0  &     0 &    2+S_d(\WBLO+\Omega)  &     0 \\
      \end{array}
    \right) \quad.
\end{equation*}

As stated in the main text, it is possible to tranform $\matCA^\phi$ into $\matCA$ of Eq.~(\ref{eq_Aout}), via 

\begin{equation}   
\label{eq_AA_phases}
    \matCA= \mathbf{R}\, \matCA^\phi\, \mathbf{R}
\end{equation}

\noindent where $\mathbf{R}$ is given by $\mathbf{R} = \mathrm{Diag}[ e^{i\phi_A},\, e^{-i\phi_A},\, e^{i\phi_B},\, e^{-i\phi_B}]$.

The phases $\phi_A$ and $\phi_B$ are determined through fitting with Eq.~(\ref{eq_Aout}) as is discussed in Sec.~\ref{supp_par_est}. Anyway, we remark that terms $A^\phi_{1,2}\equiv A_{1,2}$ and $A^\phi_{3,4} \equiv A_{3,4}$ are phase-independent as, indeed, they represent the standard heterodyne PSD. This justifies the procedure of fitting the heterodyne spectra using $\matCA^{\phi}$, before determining $\phi_A$ and $\phi_B$.

\subsection{Particle positioning}

The particle position is centered on the cavity axis by scanning along the cavity transverse plane and recoding the resulting beat note amplitude in the heterodyne detection (see for example Ref.~\cite{Ranfagni2021}). Typically, it is convenient to place the nanoparticle as close as possible to a cavity node in order to minimize heating rates associated with the lasers frequency noise~\cite{Rabl2009,Delic2020}. However, this also minimizes the amplitude of the beat notes making more difficult to obtain a stable phase reference in the post processing of the data. As a compromise, we shift the particle mean position from the node by approximately $\simeq1$\,nm. This is sufficient to obtain clean beat notes and, at the same time, is small enough to ensure negligible contribution from frequency noise. Indeed, independent measurements indicate that a $1$\,nm shift corresponds to an excess noise in detection of less than $3\%$.

\begin{figure}[!h]
\includegraphics[width=9 cm]{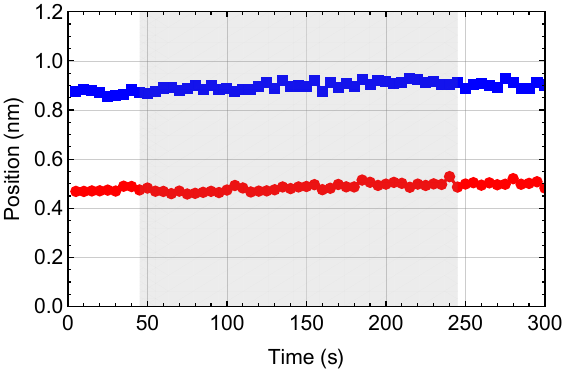}
\caption{Mean particle distance from a cavity node of field $A$ (red) and $B$ (blue). It is calculated by calibrating the rms amplitude of the two fields beat note, one of which ($A$) is stabilized with active feedback.  The shaded region corresponds to the acquisition period of the data shown in Figs.~\ref{fig_reAA_sup}
- \ref{fig_imAA_sup}.}\label{fig_ampLO_sup}
\end{figure}

In order to improve long term stability of the particle position, we also implemented a feedback loop to maintain constant beat note amplitudes. The scheme is a side-lock with a bandwidth of $\,\sim2\div3$\,Hz, acting on the piezoelectric actuator that controls the position of the tweezer lens along the cavity axis, and using as input the amplitude of the beat note of field $A$. This allows us to compensate for slow thermal drifts of the particle position during the acquisition of the heterodyne time traces. This automatically fixes the amplitude of field $B$ beat note up to small changes of the cavity free spectral range.

We show in Fig.~\ref{fig_ampLO_sup} an example of the time evolution of both beat note amplitudes with the feedback engaged. Each data point corresponds to an averaging time of $5$\,s. The amplitude is calibrated to indicate the mean particle distance from the respective closest cavity node. This distance is not the same since the two fields drive different cavity longitudinal modes, separated by two FSR. From the difference between the two distances, we deduce that the nanosphere is positioned at $\sim18\,\mu$m from the cavity center, along its axis.  

\subsection{Parameters estimation}\label{supp_par_est}

\begin{figure}[h]
\includegraphics[width=12cm]{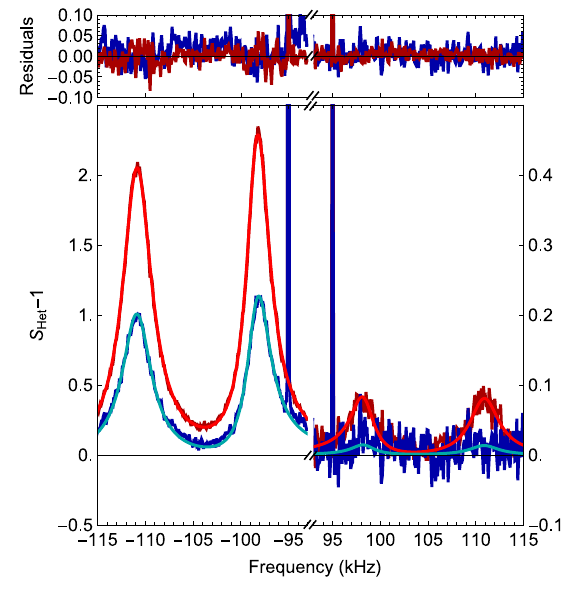}
\caption{Fit of heterodyne spectra using Eqs.~(\ref{eq_ShetA}-\ref{eq_ShetB}). The main panel shows $S_{\mathrm{het}}^{\scriptscriptstyle{A}}$ (dark red) and $S_{\mathrm{het}}^{\scriptscriptstyle{B}}$  (dark blue) along with the fitted curves for field $A$ (light red) and $B$ (light blue). The detuning were set to $\Delta_B=-\Delta_A=2\pi\,100$\,kHz. For clarity of visualization, the original frequency axis direction has been used for $S_{\mathrm{het}}^{\scriptscriptstyle{A}}$.  The top panel shows the fit residuals, following the same color pattern as in the main panel. The returned fit parameters are $(\Omega_x,\Omega_y,g_{x},g_y,\Gamma_x,\Gamma_y,\Delta_{off})/2\pi=(110.51\pm0.03,\, 98.29\pm0.04,\, 9.01\pm0.18,\, 7.50\pm0.06,\, 3.00\pm0.05,\, 2.67\pm0.05,\, 1.35\pm0.18)$\,kHz and a power ratio $\Prat=0.292\pm0.003$.
}\label{fig_het_fit_sup}
\end{figure}

The dynamical parameters and heating rates are obtained by simultaneously fitting the heterodyne spectra $S_{\mathrm{het}}^{\scriptscriptstyle{A}}(\Omega)$ and $S_{\mathrm{het}}^{\scriptscriptstyle{B}}(\Omega)$ using Eqs.~(\ref{eq_ShetA}-\ref{eq_ShetB}). An example of this type of fit is shown in Fig.~\ref{fig_het_fit_sup}, using the same dataset presented in Fig.~\ref{fig_raw_supp} and Fig.~2 of the main text. The free parameters are the mechanical frequencies $\Omega_i$, coupling rates $g_i$, heating rates $\Gamma_i$ with $i=(x,y)$, the effective power ratio $\Prat$, and a common mode detuning offset $\Delta_{off}$ that accounts for  long term drifts of the reference laser locking point. Other necessary parameters are fixed to the independently measured values, i.e., $\kappa/2\pi=58\pm0.6\,$kHz and $\eta=0.283\pm0.006$. The frequency range shown in Fig.~\ref{fig_het_fit_sup} exactly corresponds to the fitted region. Also shown are the fit residuals, which demonstrate excellent agreement between data and model. 

\begin{figure}[h!]
\includegraphics[width=13 cm]{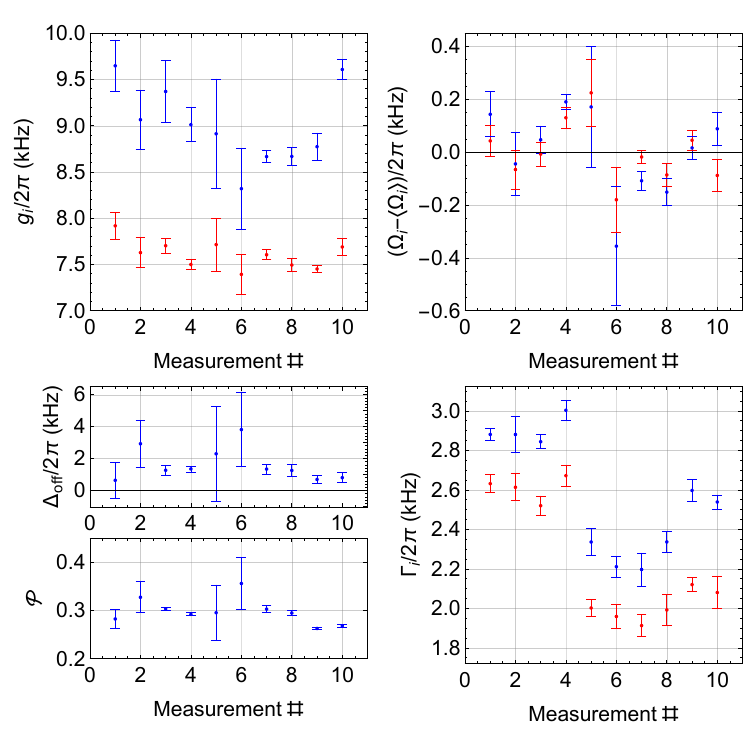}
\caption{Estimated system parameters for all datasets, obtained by fitting the heterodyne spectra. (a) Optomechanical coupling rates $g_x$ (blue) and $g_y$ (red). (b) Trap frequencies, shown as shift with respect to mean values (same color code as in (a)). (c) and (d) show respectively the common mode detuning offset $\Delta_{off}$ and power ratio $\mathcal{P}$. (e) Heating rate $\Gamma_x$ (blue) and $\Gamma_y$ (red). 
}\label{fig_all_par_sup}
\end{figure}

We summarize in Fig.~\ref{fig_all_par_sup} the parameters extracted from all the datasets discussed in the main text. The coupling rates shown in Fig.~\ref{fig_all_par_sup}\,(a) indicate that $g_x$ us more stable than $g_y$. This was expected since $g_x\propto\sin^2 \theta_p$ and $g_y\propto\sin \theta_p \cos \theta_p$, where $\theta_p$ is the angle between the cavity axis and the tweezer field polarization. In the current experiment $\theta_p\simeq45\degree$ so that $g_x$ has maximal sensitivity to any change/drift in the polarization state at the fiber output, while $g_y$ is only weakly sensitive. The estimated trap frequencies are shown in Fig.~\ref{fig_all_par_sup}\,(b) in the form of a frequency shift with respect to the mean values, $\langle \Omega_x \rangle=110.32$\,kHz and $\langle \Omega_y \rangle=98.16$\,kHz, across all measurements. The fluctuations remain significantly smaller than the modes effective linewidths which, in the worst case scenario occurring at the largest detuning ($150$\,kHz), is for both modes $\sim1$\,kHz. The detuning offset $\Delta_{off}$, shown in Fig.~\ref{fig_all_par_sup}\,(c), represents a small correction to the nominal detuning for all datasets. The power ratio $\mathcal{P}$ remains fairly constant throughout the experiment at a value of $\simeq0.3$, with only the last two datasets showing a small deviation of a few percent. The last parameters are the heating rates, which form two distinct groups reflecting a measured reduction of the chamber pressure by $\sim30\%$.

\begin{figure}[h!]
\includegraphics[width=\textwidth]{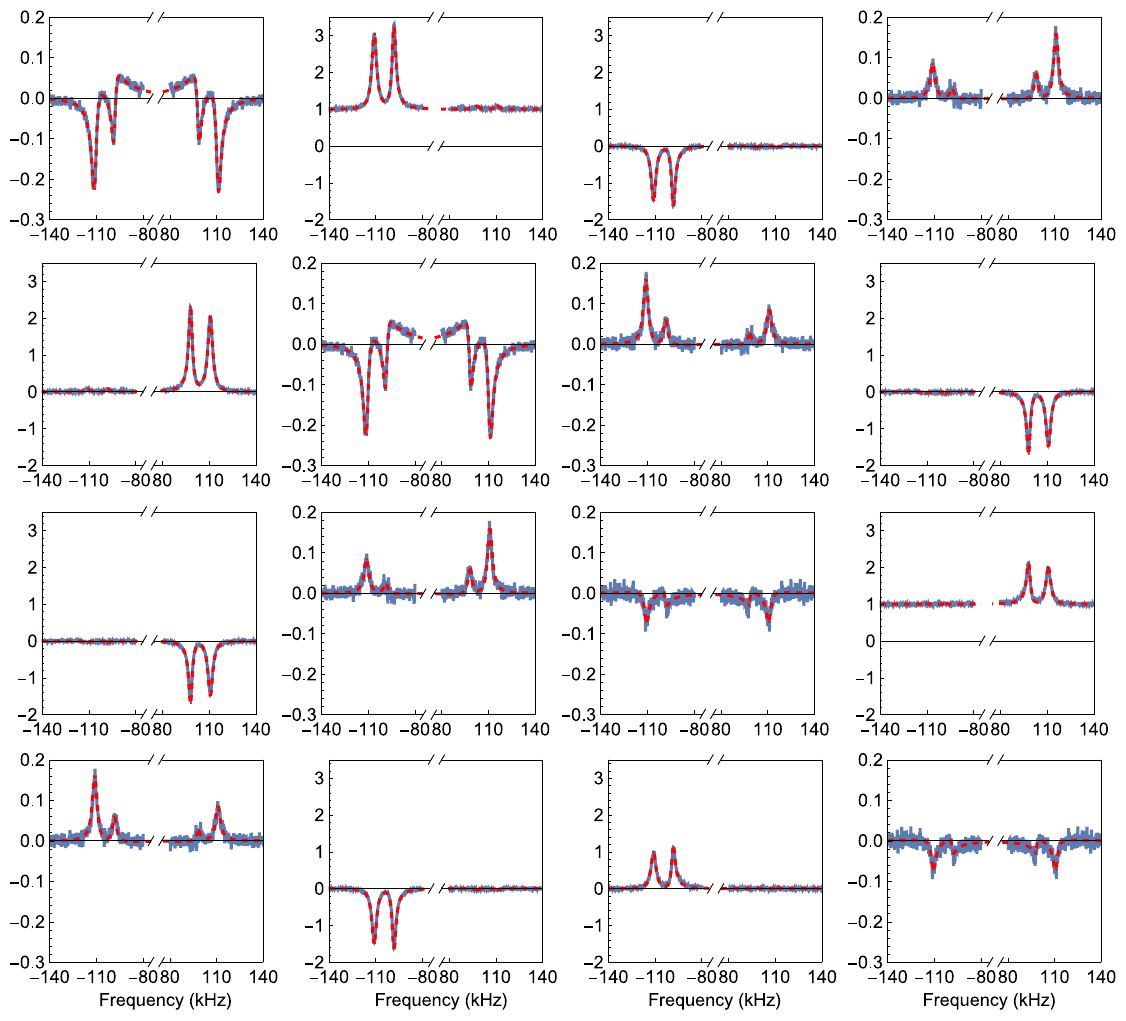}
\caption{Experimentally determined spectral correlation matrix $\matCA$. Each panel shows the real part of the corresponding matrix element of $\matCA$. The experimental data (blue) are compared to the model (dashed-red) defined in Eq.~(\ref{eq_Aout}).  The only elements carrying a shot noise contribution are the $\matCA^\phi_{1,2}$  and $\matCA^\phi_{3,4}$ corresponding to correlators $\langle a^{\alpha}\,a^{\alpha,\dagger}\rangle$.}
\label{fig_reAA_sup}
\end{figure}

\begin{figure}[h!]
\includegraphics[width=\textwidth]{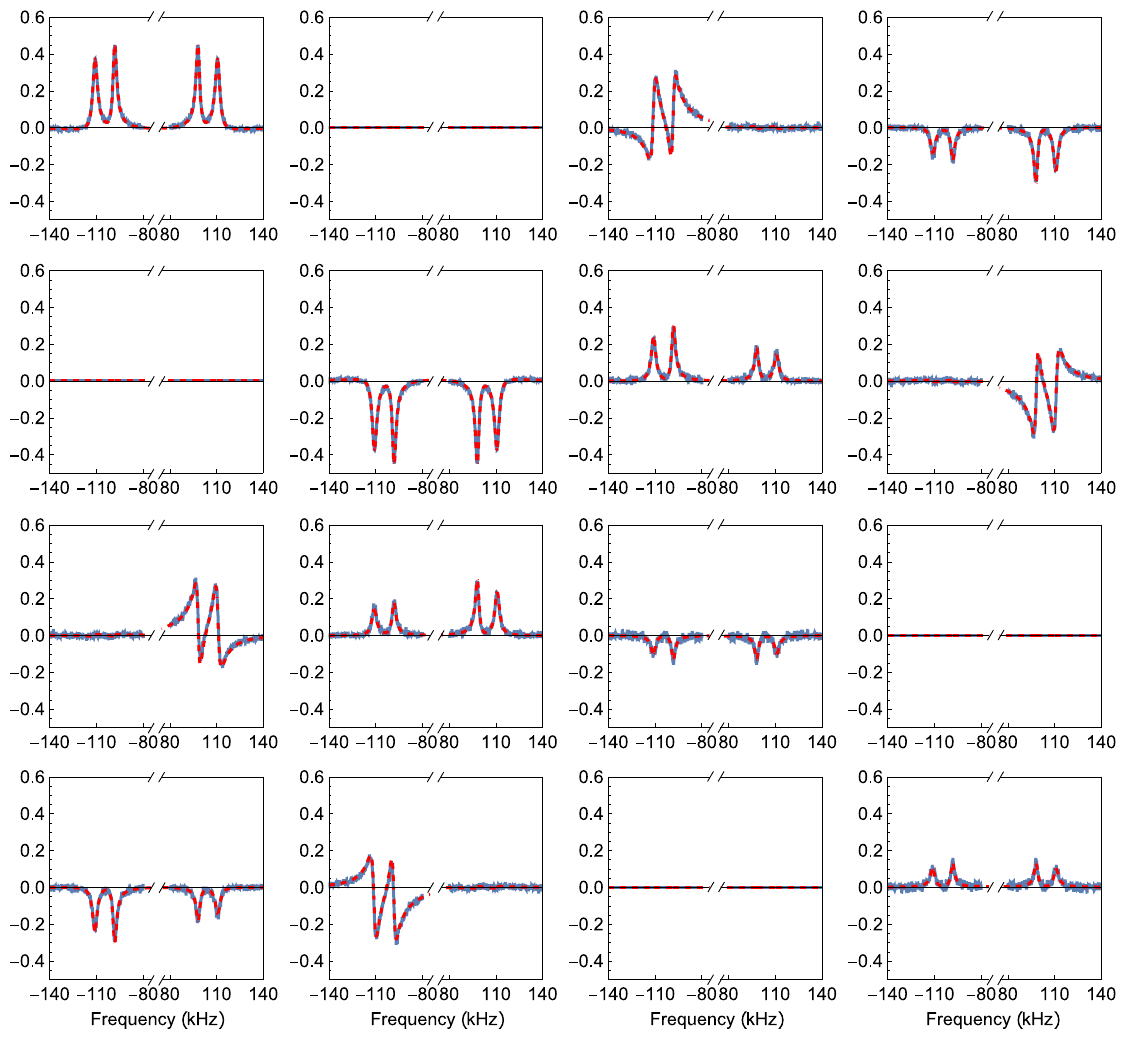}
\caption{Experimentally determined spectral correlation matrix $\matCA$. Each panel shows the imaginary part of the corresponding matrix element of $\matCA$. The experimental data (blue) are compared to the model (dashed-red) defined in Eq.~(\ref{eq_Aout}).  Off-diagonal elements of the two fields sub-matrices are vanishing since they are defined to be real.
}\label{fig_imAA_sup}
\end{figure}

The final step for the determination of the experimental $\matCA$ matrix is the estimation of the phases $\phi_A$ and $\phi_B$. To this end, we transform the experimental $\matCA^\phi$ matrix according to Eq.~(\ref{eq_AA_phases}) and compare it to Eq.~(\ref{eq_Aout}). More precisely, we simultaneously fit all the real parts of the matrix elements, using only $\phi_A$ and $\phi_B$ as free parameters. All other parameters are fixed to the values obtained by fitting $S_{\mathrm{het}}^{\scriptscriptstyle{\alpha}}$.  The result of this second fit, for the same data as in Fig.~\ref{fig_het_fit_sup}, is shown in Fig.~\ref{fig_reAA_sup} and Fig.~\ref{fig_imAA_sup}  where the real and imaginary parts are plotted, respectively.

\newpage

All spectral elements in Figs.~\ref{fig_reAA_sup}-\ref{fig_imAA_sup} show fantastic agreement between experimental data and modeling. This clearly demonstrates how well the model captures the physics of the experiment, since fitting the heterodyne PSDs enables the prediction of all cross-correlation terms in $\matCA$ up to two constant phase factors.

\subsubsection{Spectrum of the z mode}
\label{Sec_Zexp}

\begin{figure}[h]
\includegraphics[width=8cm]{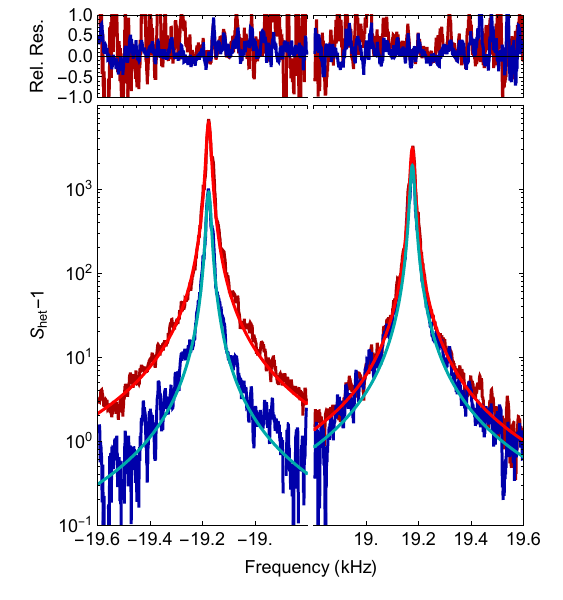}
\caption{Fit of heterodyne spectra using Eq.~(\ref{eq_ShetZ}) around the resonance of the particle motion along the tweezer axis (\textit{z}). The main panel shows $S_{\mathrm{het}}^{\scriptscriptstyle{A}}$ (dark red) and $S_{\mathrm{het}}^{\scriptscriptstyle{B}}$  (dark blue) along with the fitted curves for field $A$ (light red) and $B$ (light blue). The top panel shows the fit residuals, following the same color pattern as in the main panel. The effective linewidth is $\GeffZ/2\pi=14.5$, and the fit parameters are $(\Omega_z,\gAZ,\gBZ,\GZ)/2\pi=(19.25,\,2.25,\,1.22,\,32.6)$\,kHz.
}\label{fig_het_Z_supp}
\end{figure}

A typical example of the spectral region around the resonance frequency of the $z$ mode is shown in Fig.~\ref{fig_het_Z_supp}, together with a fit using Eq. (\ref{eq_ShetZ}). The parameters obtained from the fit are  $\WZ/2\pi = 19.25\, \mathrm{kHz}$, $\GZ/2\pi = 32.6\, \mathrm{kHz}$, $\gAZ/2\pi = 2.25\, \mathrm{kHz}$ and $\gBZ/2\pi = 1.22 \,\mathrm{kHz}$. 

The coupling rates remain stable within few percent, and the heating rate within $10\%$, during the whole experiment. 
These values, inserted into Eq. (\ref{eq_sigmaadd}), allow us to quantify the contribution of the $z$ mode, mediated by the cavity field, to the variance of the bright mode. For example, for the optimal detuning, we obtain $\, \langle \xb\xb \rangle ^{\scriptscriptstyle{\mathrm{add}}}= 0.22$, and  $\, \langle \pb\pb \rangle ^{\scriptscriptstyle{\mathrm{add}}}= 0.007$. 

\newpage

\section{Optomechanical entanglement}\label{Sec_entanglement}

\begin{figure}[h!]
\includegraphics[width=\textwidth]{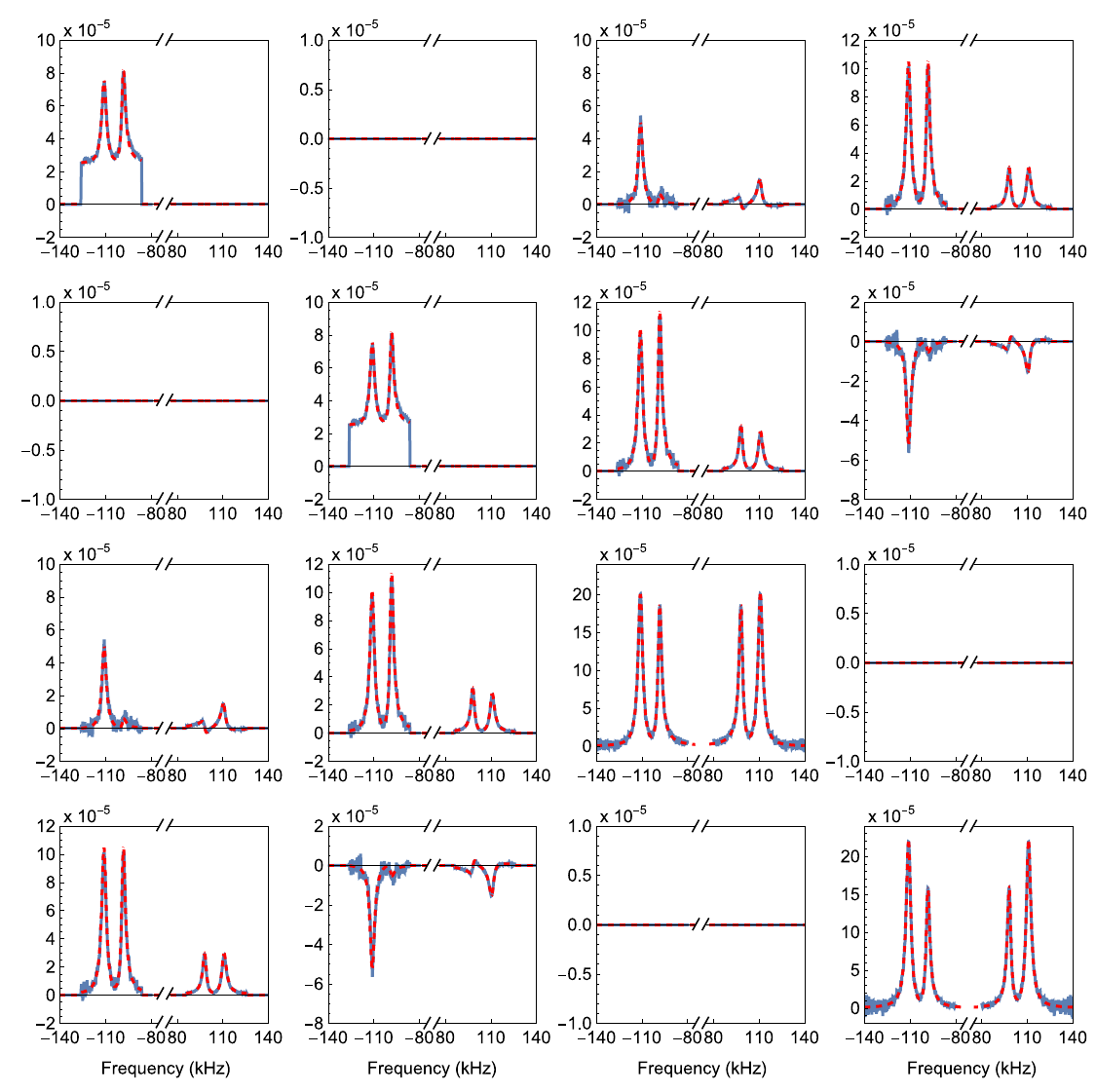}
\caption{Spectral elements of the matrix $\bar{\mathbf{S}}^{\scriptscriptstyle{\mathrm{m}}}$ considering $\Gammaxi/2\pi=40$\,kHz. Elements $\bar{S}_{3,3}$ and $\bar{S}_{4,4}$ represent the symmetrized position and momentum spectra, respectively. The experimentally reconstructed spectra (blue) are compared to the model (dashed-red) according to Eq.~(\ref{eq_SBb}).
}\label{fig_QX_sup}
\end{figure}

Once the spectral correlation matrix $\matCA$ has been obtained and the system parameters have been estimated, the covariance matrix $\mathbf{V}^{\scriptscriptstyle{\xi B}}$ can be determined through Eqs.~(\ref{eq_SBb}-\ref{eq_VBb}). As stated in the main text, we define $\xi(\omega) = 1/\sqrt{2 \pi \Gammaxi}$ for $(-\Wb-\Gammaxi/2) < \omega <  (-\Wb+\Gammaxi/2)$, and $\xi(\Omega)=0$ elsewhere. This choice leaves $\Gammaxi$ as the only free parameter, which can be optimized. We show in Fig.~\ref{fig_QX_sup} the result of this calculation right at the step before integration, i.e., after having calculated $\bar{\mathbf{S}}^{\scriptscriptstyle{\mathrm{m}}}=0.5\, \mathrm{Re} \left[ \mathbf{S}^{\scriptscriptstyle{\mathrm{m}}}(\omega) + \mathbf{S}^{\scriptscriptstyle{\mathrm{m}} \scriptstyle{\mathrm{T}}} (\omega)  \right]$, for the optimal $\Gammaxi/2\pi=40$\,kHz. Here we consider only the 4x4 matrix for the variables of interest, that is $\mathbf{q}^{\scriptscriptstyle{\xi}} = (Q_{\xi,\scriptscriptstyle{A}},\,P_{\xi,\scriptscriptstyle{A}},\,\xb,\,\pb)^{\mathrm{T}}$. The data shown are the same used for Figs.~\ref{fig_het_fit_sup},\ref{fig_reAA_sup},\ref{fig_imAA_sup}. Even at this stage, we observe a great agreement between the experimental data and the model.

The last step to obtain the covariance matrix $\mathbf{V}$ is direct integration. The adopted definition of the filter function $\xi(\omega)$ has the clear advantage that the integrals receive non-zero contributions only within a frequency bandwidth where the signal-to-noise ratio (SNR) is maximal for all matrix elements associated with field $B$. However, this is not the case for the position and momentum spectra, which require integration from $-\infty$ to $\infty$. This can be problematic, since the SNR drops dramatically on the tails of the resonances of the bright mode. To address this issue, all elements of $\mathbf{V}$, except $\mathrm{V}_{3,3}$ and $\mathrm{V}_{4,4}$, are obtained by direct summation of the spectra. For these last two elements, direct integration is performed only over a bandwidth of $\sim40$\,kHz centered on the bright mode frequency $\Omegab$, and is then corrected by adding a numerical integration of the model outside this region. This correction is typically on the order of $\sim10\%$ of the total integral value.

In all cases, before calculating the values of $\num$ reported in this article, we added the contribution of the $z$ mode to $\langle \xb\xb \rangle$ and $\langle \pb\pb \rangle$, calculated as described in Sections \ref{Sec_axial} and \ref{Sec_Zexp}, to the appropriate diagonal entries of the covariance matrix. The increase of $\num$ due to these higher variances is about $0.015 \div 0.03$, comparable to the experimental uncertainty.

For completeness, we show in Figs.~\ref{fig_cov_sup}-\ref{fig_cov2_sup} (at the end of this document) both the 2D portrayal of the covariance matrix $\matV$ and the corresponding numerical values for all the datasets discussed in the main text, i.e., for all the data points shown in its Fig.~4. For each matrix, the nominal detuning and the mode width $\Gammaxi$ are listed. Note that the same color scale is used in all 2D representations.

\subsection{Intracavity entanglement}
\label{Sec_intracavity}

The parameters obtained from the fits to the heterodyne spectra are used as inputs to the model described in Section \ref{Sec_IIa} to determine the covariance matrix $\mathbf{V}^{\mathrm{bd}}$ of the cavity optomechanical system.

When the $z$ mode is included in the model, extending the description from the two-dimensional transverse plane, the variance of the bright mode increases, in agreement with the predictions of Section \ref{Sec_axial}. More importantly, unlike the case of the propagating optical mode, the intracavity covariance matrix now incorporates the influence of the $z$ mode also in the $B$ field and in the optomechanical correlations. These contributions to the optical subsystem vanish in $\mathbf{V}^{\scriptscriptstyle{\xi B}}$ because the filter $\xi(\omega)$ removes the spectral region around the $z$-mode resonance. As a result of this additional noise, the two-dimensional bright mode is no longer entangled with the cavity field. Nevertheless, optomechanical entanglement is partially recovered by considering a slightly tilted direction of motion, characterized by the angle $\thz$, which enhances the optomechanical correlations in the spectral region of the $z$ mode.  An example of the matrix $\mathbf{V}^{\mathrm{bd}}$ calculated for an optimal $\varphi$ is given below:
\begin{equation*}
\mathbf{V}^{\mathrm{bd}} =
\begin{tikzpicture}[baseline=(M.center)]
\matrix (M) [matrix of math nodes,left delimiter=(,right delimiter=),nodes={anchor=east}] {
        13.1  &  2.55  & -6.61  &  1.36  & 32.35  & -0.99 & -0.31 &   0.   & -361    &  20.67   \\
         2.55 &  1.97  & -1.42  &  0.56  &  7.57  &  1.22 & -0.07 &   0.03 & -78.8   & -44.66   \\
        -6.61 & -1.42  &  4.61  & -0.76  & -17.7  &  0.75 &  0.17 &  0.04  & 196     & -11.26   \\
         1.36 &  0.56  & -0.76  &  1.33  &  4.31  &  0.92 & -0.01 &  0.02  & -42.9   & -24.27   \\
         32.4 &  7.57  & -17.7  &  4.31  &  88.7  &  0.   & -1.27 &  0.36  & -964    &   3.43   \\
        -0.99 &  1.22  &  0.75  &  0.92  &  0.    &  7.76 & -0.35 & -0.08  & -6.63   & -171     \\
        -0.31 & -0.07  &  0.17  & -0.01  & -1.27  & -0.35 &  5.30 &  0.    &  7.04   & -0.44    \\
         0.   &  0.03  &  0.04  &  0.02  &  0.36  & -0.08 &  0.   &  5.22  &  0.08   &  1.27    \\
       -361   & -78.8  &  196   & -42.9  & -964   & -0.63 & -6.63 &  7.04  & 11\,133 &   0.     \\
        20.7  & -44.7  & -11.3  & -24.3  &   3.43 & -171  & -0.44 &  1.27  &  0.     & 10\,691  \\
};
\draw[blue,dashed,thick,rounded corners]
(M-3-3.north west) rectangle (M-6-6.south east);
\end{tikzpicture}
\end{equation*}

\noindent The blue square highlights the relevant block giving $\mathbf{V}^{\mathrm{Bb}}$. The parameters used to calculate it are $(\OmegaX,\OmegaY,\gX,\gY,\GX,\GY,\Delta_A,\Delta_B)/2\pi=(110.5,\,98.4,\,8.9,\,7.7,\,2.3,\,2.0,\,-137.8,\,137.4)$\,kHz and $\Prat=0.292$. They correspond to one of the experimentally estimated parameters sets. The optimal tilt angle is, in this case, $\varphi=2.0\degree$.

\begin{figure}[h!]
\includegraphics[width=\textwidth]{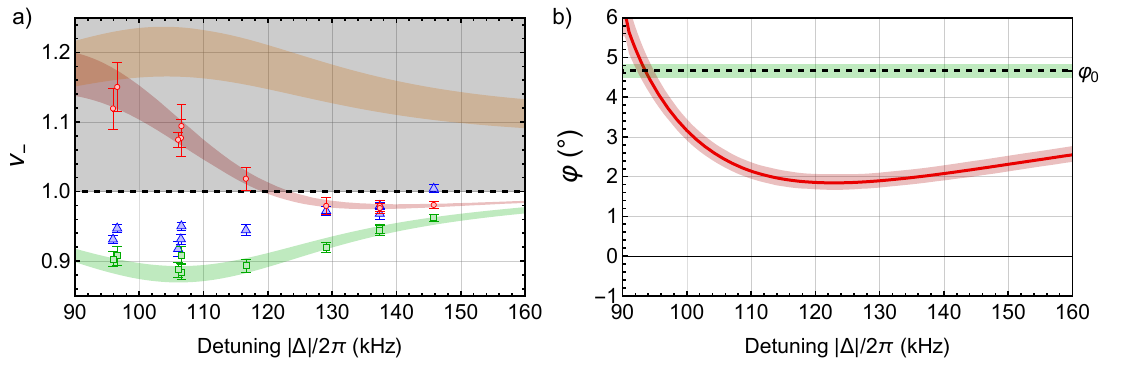}
\caption{ \textbf{a)} Symplectic eigenvalue $\num$ versus cavity detuning, as in Fig.~4 of the main text. Experimental data points, obtained through Eq.~(\ref{eq_VBb}), are shown by blue markers with error bars. Green markers are extracted through the optomechanical model (Eq.~(\ref{eq_Veta_a})), using the parameters that best fit the heterodyne spectra. In these measurements, we set $-\Delta_A\simeq\Delta_B=|\Delta|$. Intracavity entanglement, red data points, is inferred from the covariance matrix calculated from Eq.~(\ref{eq_covvlyap}), where for each point the tilt angle $\phi$ has been optimized. The green and red shaded regions show extremal model predictions when considering the propagating and cavity mode respectively. The orange shaded region has been calculated using the fix tilt angle $\thz_0$. \textbf{b)} Optimal tilt angle $\thz$ (red) that minimizes the eigenvalue $\num$, thus allowing to infer intracavity entanglement. The values of $\thz_0$ is also shown as reference (black-dashed line). Shaded regions reflect the variability of the system parameters illustrated in Fig.~\ref{fig_all_par_sup}.
}\label{fig_detuning_supp}
\end{figure}

We show in Fig.~\ref{fig_detuning_supp}\,a), the symplectic eigenvalue $\num$ as a function of the detuning, as in Fig.~4 of the main text, but with an extended vertical scale that allows us to fully show the behavior $\num$ characterizing the intracavity system. The latter has been calculated by optimizing the tilt angle $\thz$ for each detuning and set of measured parameters. The resulting optimal $\thz$ is shown in Fig.~\ref{fig_detuning_supp}\,b), along with $\thz_0=4.66\degree\pm0.17\degree$. For this fixed tilt angle, the calculated eigenvalue (shaded orange region in Fig.~\ref{fig_detuning_supp}\,a) always remains above the separability threshold.

\section{Error estimation}

Statistical errors are always evaluated as sample statistic. To this end, the time traces of each dataset are divided into shorter blocks, typically between 5 and 10. The full analysis, described in Sec.~\ref{sec_Amatrix}-\ref{Sec_entanglement}, is then performed for each block independently, yielding at each step an ensemble of values for all relevant parameters from the different blocks. 

The systematic uncertainty is calculated taking into account the uncertainty of the independently measured parameters, i.e., the cavity full linewidth $\kappa$ and the detection efficiency $\eta$. At this purpose, the analysis is carried out by assigning three possible values to both $\kappa$ and $\eta$: their mean value, as well as the upper and lower bounds defined by their respective uncertainties. All possible combinations of these choices are considered, resulting in a total of nine configurations for each dataset. We have taken very conservative values of the uncertainties $\Delta\kappa=1$\,kHz and $\Delta\eta=0.025$. The systematic error on $\num$ quoted in the main text reflects the observed maximal values at the end of this procedure.

\subsection{Effect of the finite number of measurements}

A covariance matrix reconstructed from a finite number of quadrature measurements provides a reliable estimator of the system covariance matrix, i.e.,
\begin{equation}
    \mathrm{E}\left[\frac{1}{N}\sum_{n=1,  N}\, u^{(n)}_i u^{(n)}_j\right] = \langle u_i u_j\rangle
\end{equation}
where $u_i^{(n)}$ denotes the n-th measurement of the quadrature $u_i$. This is not true for $\num$, which depends nonlinearly on the elements of $\mathbf{V}$. Therefore, any estimate of $\num$ based on a finite data sample is biased, and its accuracy must therefore be carefully assessed.

\begin{figure}[h!]
\includegraphics[width=10 cm]{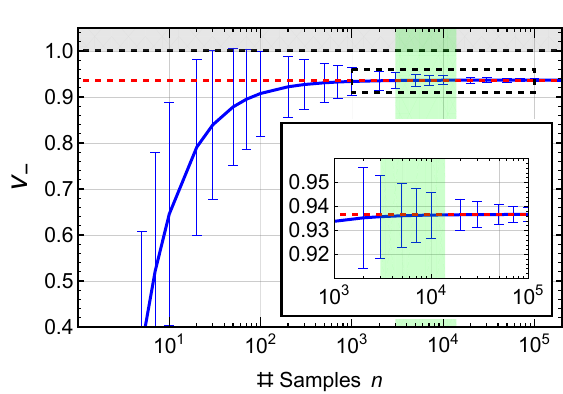}
\caption{Simulated estimate of $\num$ from a known covariance matrix, as a function of the number of independent samples $n$ used for each individual estimation of $\num$. The true value of $\num = 0.937$ is also shown (red-dashed line). For small $n$ a significant bias is present. However, for values of $n$ in the range of our experiment (green shaded region) the bias becomes negligible. The inset shows an expanded view of the region most relevant to the experimental results.   
}\label{fig_bias_supp}
\end{figure}

\begin{figure}[h!]
\includegraphics[width=\textwidth]{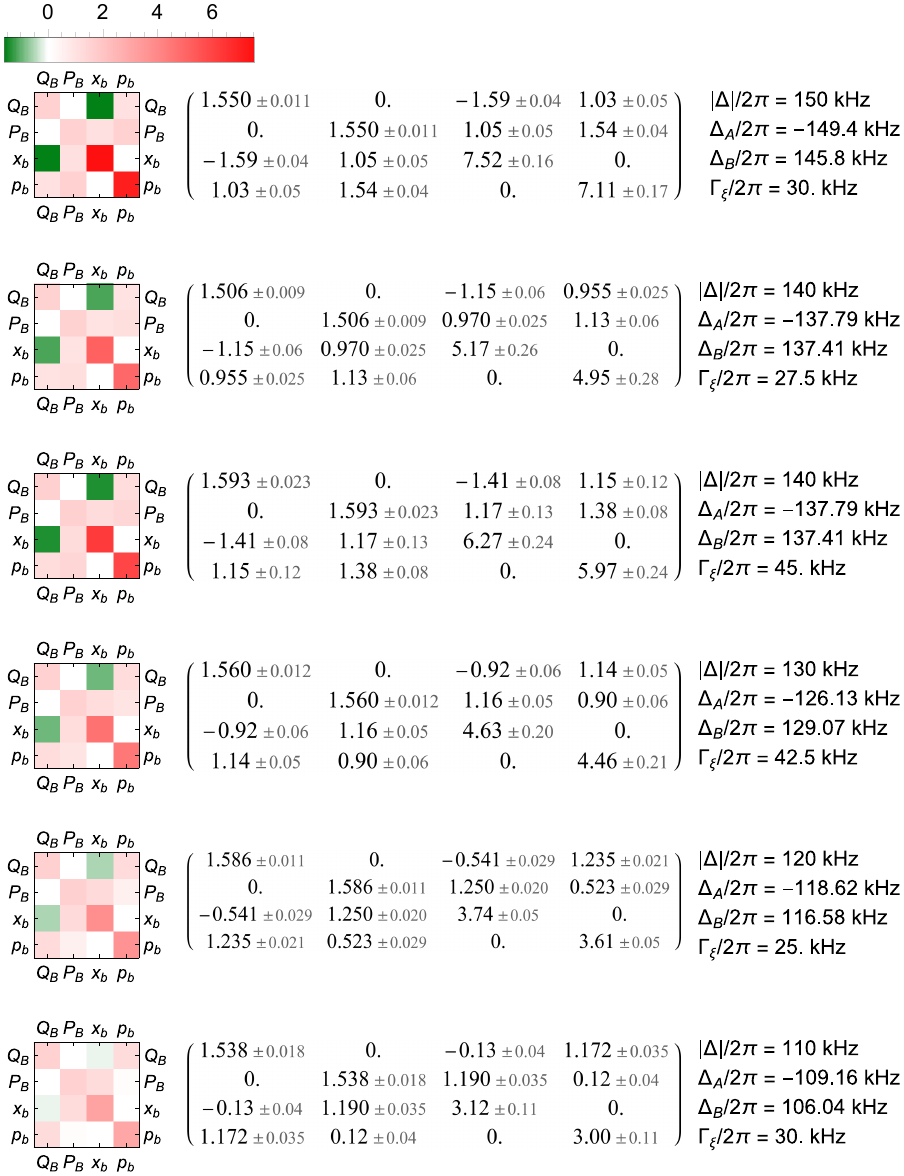}
\caption{2D portrayal of the covariance matrix $\matV$ fully characterizing the optical-mechanical Gaussian state, and corresponding numerical values (First part).
}\label{fig_cov_sup}
\end{figure}

\begin{figure}[h!]
\includegraphics[width=\textwidth]{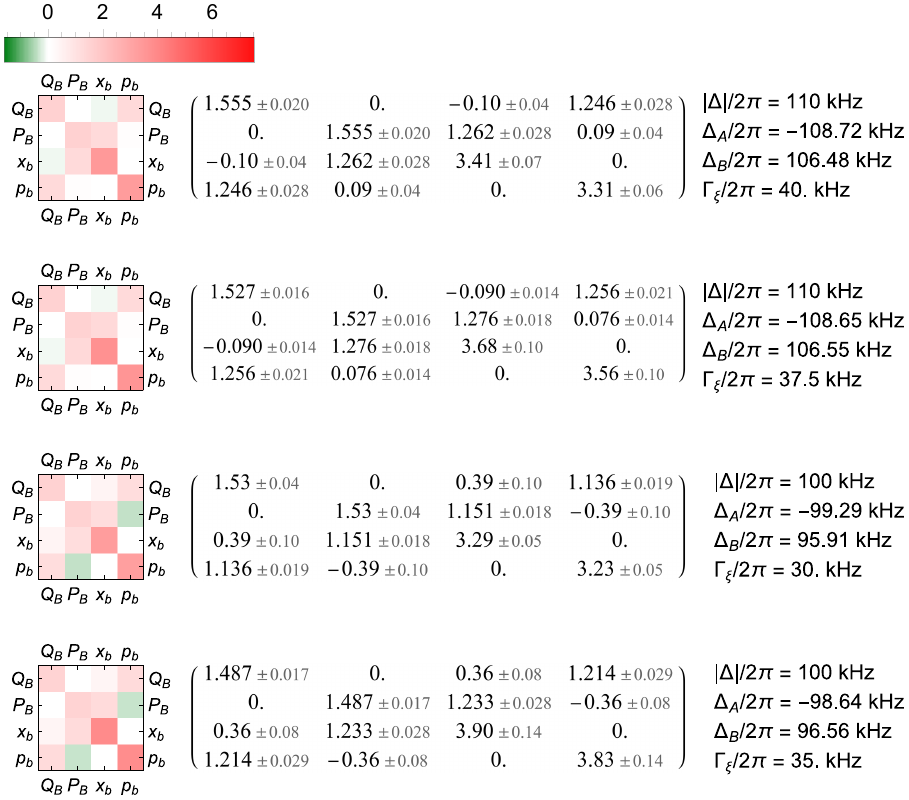}
\caption{2D portrayal of the covariance matrix $\matV$ fully characterizing the optical-mechanical Gaussian state, and corresponding numerical values (Second part).
}\label{fig_cov2_sup}
\end{figure}

We have performed a numerical simulation by generating $10^7$ sets of four Gaussian stochastic variables $\mathbf{u}$, with a probability distribution $\mathrm{P}(\mathbf{u})$ generated by a typical covariance matrix $\mathbf{V}$ of our system, i.e.,  
\begin{equation}
    \mathrm{P}(\mathbf{u})=\frac{1}{(2\pi)^2\,\sqrt{\mathrm{Det}[\mathbf{V}]}}\,\exp\left(-\frac{1}{2}\, \mathbf{u}^{\mathrm{T}}\, \mathbf{V}^{-1} \mathbf{u}\right)  \, .
\end{equation}
We have then reconstructed the covariance matrix from subsets of $n$ samples of $\mathbf{u}$, computed $\num$ for each reconstructed matrix, and evaluated the corresponding mean value and standard deviation of $\num$ as a function of $n$. The results, shown in Fig.~\ref{fig_bias_supp}, clearly indicate that for small $n$ the estimator of $\num$ significantly deviates from its true value, calculated from the original $\mathbf{V}$. Therefore, the number of measurements employed to reconstruct each covariance matrix should be sufficiently large to ensure that this deviation remains below the experimental uncertainty. The drawback is that, over longer timescales, maintaining sufficient stability of the system parameters becomes increasingly challenging, resulting in a smearing of the measured correlations between the two subsystems. Consequently, finding an appropriate compromise is a key experimental requirement.    

In order to use the numerical results to assess the accuracy of our experimental estimate of $\num$, we have to evaluate the effective number of independent quadrature measurements used to compute each value of $\num$. An upper bound is given by $\,\mathrm{T}_{\scriptscriptstyle{\mathrm{M}}}\,\Gammaxi/2\pi$, where $\mathrm{T}_{\scriptscriptstyle{\mathrm{M}}}$ is the total measurement time employed to reconstruct the covariance matrix (this is a correct estimate for stationary white noise). A lower bound can be extracted directly from the experiment by repeating the reconstruction of $\mathbf{V}$ several times with the same $\mathrm{T}_{\scriptscriptstyle{\mathrm{M}}}$, evaluating the mean value $\mu \left[ \mathrm{V}_{ii} \right]$ and the standard deviation $\sigma \left[ \mathrm{V}_{ii} \right]$ of $ \mathrm{V}_{ii}$, and defining an effective number of independent measurements as $n_{\mathrm{eff}} = \left(\mu \left[ \mathrm{V}_{ii} \right]/\sigma \left[ \mathrm{V}_{ii} \right] \right)^2$. This definition follows from the standard $1/\sqrt{n}$ scaling of statistical fluctuations for independent measurements. 

This conservative estimate of $n_{\mathrm{eff}}$ ranges between 3000 and 14000 ($\mathrm{T}_{\scriptscriptstyle{\mathrm{M}}}$ between $10$\,s and $40$\,s) for all the measurements reported. As shown in Fig.~\ref{fig_bias_supp}, the corresponding bias is negligible. We further note that the statistical uncertainty obtained from the numerical simulation is in good agreement with the experimental uncertainties.  

\newpage

\bibliographystyle{nicebib}
\bibliography{database}